\documentclass[twocolumn,showpacs,preprintnumbers,amsmath,amssymb,aps]{revtex4}
\usepackage{graphicx}
\usepackage{amsmath}

\begin{document}
\title{General Theory of the Quantum Kicked Rotator. I}
\author{Tao Ma}
\affiliation{Department of Modern Physics, University of Science and
Technology of China, Hefei, PRC}
\email{taomascience@gmail.com}
\date{\today}
\begin{abstract}
This is the first of a series of two papers. We discuss some basic
problems of the quantum kicked rotator (QKR) and review some
important results in the literature. We point out the flaws in the
inverse Cayley transform method to prove dynamic localization. When
$\tau/2\pi$, where $\tau$ is the kick period, is very close to a
rational number, the localization length is larger than the typical
localization length. We analytically prove anomalous localization
and confirm it by numerical calculations. We point out open problems
that need further work.
\end{abstract}
\maketitle
\section{\label{sec:level1}Physical explanations of Quantum kicked rotator}
Nearly thirty years ago, QKR is first studied by G. Casati, B. V.
Chirikov, F. M. Izraelev and J. Ford \cite{Casati1979}. They
discovered by numerical calculations when the kick period $\tau$ is
the product of an irrational number and $2\pi$, the rotator
localizes in the momentum space. Later, S. Fishman, D. R. Grempel
and R. E. Prange explained the localization by transforming QKR into
an Anderson localization problem \cite{Fishman1982}. It does not
seem necessary to discuss the basic problems of QKR again. But many
problems are impossible to be solved by present methods. The paper
is both a review and a problem list.

The Hamiltonian of QKR is defined as \cite{Casati1979}
\begin{equation}
\begin{split}
H&=H_{0}+V(\theta) \sum_{n=1}^\infty \delta(t-n\tau) \\
&=-\frac{1}{2}\hbar^{2}\frac{\partial^2}{\partial\theta^{2}}-k
\cos\theta\sum_{n=1}^\infty \delta(t-n\tau),
\end{split}
\end{equation}
which describes a particle restricted to a ring with free
Hamiltonian
$H_{0}=-\frac{1}{2}\hbar^{2}\frac{\partial^2}{\partial\theta^{2}}$,
and periodically kicked by a homogeneous electric field parallel
with the ring plane. The parameters $\tau$ is the kick period and
$k$ is the kick strength. In QKR, the interaction between the
electric field and the rotator is $V(\theta)=-k \cos\theta$. But in
this paper sometimes we discuss a more general form of the
interaction $V(\theta)$. In the basis $\{ |m \rangle
=\frac{1}{\sqrt{2\pi}} e^{-i m \theta}\}$, the Hamiltonian can be
written as
\begin{equation}
\begin{split}
H=&\sum_{m=-\infty}^\infty \frac{m^2}{2}|m \rangle \langle m|-
\sum_{n=1}^\infty \sum_{m=-\infty}^\infty \\
& k(\frac{1}{2} |m \rangle \langle m+1|+\frac{1}{2}|m+1\rangle
\langle m|)\delta(t-n \tau).
\end{split}
\end{equation}

The following kick system has the same Floquet operator as the
Hamiltonian in Eq. $(2)$.
\begin{equation}
\begin{split}
H=&-\sum_{m=-\infty}^\infty (\frac{1}{2} |m\rangle \langle
m+1|+\frac{1}{2} |m+1\rangle \langle m|)+\\
&\sum_{n=1}^\infty \sum_{m=-\infty}^\infty \tau \frac{m^2}{2}|m
\rangle \langle m| \delta(t-n k).
\end{split}
\end{equation}

The physical explanation of the Hamiltonian in Eq. $(3)$ is a
particle on a one dimensional lattice. $|m \rangle$ is explained as
the $m-$th Wannier state or $m-$th site. The term $
-\sum_{m=-\infty}^\infty (\frac{1}{2} |m\rangle \langle
m+1|+\frac{1}{2} |m+1\rangle \langle m|) $ is the hopping matrix or
the kinetic energy. The particle is periodically delta-kicked by a
harmonic potential $\sum_{n=1}^\infty \sum_{m=-\infty}^\infty \tau
\frac{m^2}{2}|m \rangle \langle m| \delta(t-nk)$. Now $k$ is the
kick period or the free diffusion time of the particle, and the
potential is defined as $\tau \frac{m^2}{2}$.

Eq. $(3)$ is discovered when we solve the quantum kicked linear
rotator (QKLR) \cite{Grempel1982, TaoMa2007Toeplitz}. QKLR always
localizes except when $\tau=2\pi \times Integer$, where $Integer$ is
an integer. There is Bloch oscillation phenomenon when a lattice is
put in a homogeneous electric potential. It turns out QKLR
localization is a general Bloch oscillation. QKLR and the Bloch
oscillation are the linear Toeplitz system in two different
representations: the rotator representation and the site
representation. QKR can also be explained in the site
representation.

The site representation of QKR or Eq. $(3)$ can be another
implementation in the laboratory compared with the usual
implementation Eq. $(2)$ \cite{Moore1994, Moore1995}. The classical
correspondence of the site explanation is a periodically delta
kicked classical random walker on a lattice by the harmonic
potential (CKRW). In the rotator representation, the classical
correspondence is the standard map. We think CKRW is also very
interesting just as the standard map. For example, does CKRW
localize when the kick period $\tau$ is the product of an irrational
number and $2\pi$? Gong \textit{et al} independently gave an
equation similar to Eq. $(3)$ and proposed QKR can be implemented as
a Heisenberg spin chain subjected to a parabolic kicking magnetic
field \cite{Gong20071, Gong20072}. See \cite{Gong20072} for detailed
information.

The general kick system can be defined as
\begin{equation}
H=H_{0}+V \sum_{n=1}^\infty \delta(t-n\tau).
\end{equation}
The Floquet operator is $F=e^{-\frac{i}{\hbar}V}e^{-\frac{i}{\hbar}
H_{0} \tau}$. We can always treat $V$ as the unperturbed Hamiltonian
and $H_0$ the perturbation just we reexplain Eq. $(2)$ as Eq. $(3)$.
The essence of delta kicked system including QKR is two different
Hamiltonians acting on the Hilbert space in turn.

The following Hamiltonian has the same Floquet operator of Eq.
$(1)$.
\begin{multline}
H(t)=\begin{cases}
-\frac{1}{2}\hbar^{2}\frac{\partial^2}{\partial\theta^{2}} & \text{}\ 0 \leq t \emph{Mod}(\tau+k) \leq \tau \\
\cos\theta & \text{} \tau \leq t \emph{Mod} (\tau+k) \leq \tau+k,
\end{cases}
\end{multline}
where $t \emph{Mod} (\tau+k)$ is the product of the fraction part of
$\frac{t}{\tau+k}$ and $\tau+k$.

Apparently $n$ different Hamiltonians $\{H_0,H_1,\cdots, H_{n-1}\}$
can act on the wave function in turn. Then the Floquet operator is
just $F=e^{-\frac{i}{\hbar}H_{n-1} \tau_{n-1}} \cdots
e^{-\frac{i}{\hbar}H_1 \tau_{1}}e^{-\frac{i}{\hbar} H_0 \tau_{0}}$.
The above explanation also applies classically.

In a laboratory it is impossible to implement Eq. $(5)$ because
$H_0$ always exists. So the experimental implementation is we
increase the interaction strength and decrease the interaction time.
When the interaction strength is very large and the interaction time
is very small, it is almost a delta function.

When $M$ is very large the following system is the experimental
implementation of Eq. $(1)$.
\begin{equation}
H(t)=\begin{cases}
-\frac{1}{2}\hbar^{2}\frac{\partial^2}{\partial\theta^{2}} &
\text{}\ 0 \leq t
\emph{Mod}(\tau+\frac{k}{M}) \leq \tau \\
H_0+M \cos\theta & \text{} \tau \leq t \emph{Mod} (\tau+\frac{k}{M})
\leq \tau+\frac{k}{M},
\end{cases}
\end{equation}
or
\begin{equation}
H(t)=\begin{cases}
-\frac{1}{2}\hbar^{2}\frac{\partial^2}{\partial\theta^{2}} &
\text{}\ 0 \leq t
\emph{Mod}(\tau+\frac{1}{M}) \leq \tau \\
H_0+k M \cos\theta & \text{} \tau \leq t \emph{Mod}
(\tau+\frac{1}{M}) \leq \tau+\frac{1}{M}.
\end{cases}
\end{equation}

\section{\label{sec:level1}Flaws of the inverse Cayley transform method}
We study the Schrodinger Equation
\begin{equation}
i \hbar \frac {\partial} {\partial t} \psi (\theta,t)= H \psi (\theta,t).
\end{equation}
The dynamics of a free evolution of period $\tau$ and a kick is
described by the Floquet operator $F$.
\begin{equation}
F=e^{-\frac{i}{\hbar}V(\theta)}e^{-\frac{i}{\hbar} H_{0} \tau}
\end{equation}
The matrix elements of $F$ are given by
\begin{equation}
F_{nm}=\langle n|F|m \rangle=\exp(-i\hbar\tau \frac{m^2}{2}) i^{m-n}
J_{n-m}(\frac{k}{\hbar}).
\end{equation}
We define one period as a free evolution of period $\tau$ and a
kick, so after $n$ periods, the QKR wave function is
\begin{equation}
\psi(n\tau)=F^{n}\psi(0),
\end{equation}
where the unitary operator $U(n)=F^{n}$ maps the initial state to
the state at the time $n\tau$.
\subsection{\label{sec:level2}Meaning of dynamic localization}

Now we ask the fundamental problem of QKR. If the rotator is
initially in the ground state, will its energy $\sum _{n=-\infty
}^{\infty } \frac{1}{2} \left| c_{n} \right|^{2} n^{2}$ increase to
infinity when the time runs to infinity? This problem can also be
described in another way, such as the wave function $\sum
_{n=-\infty }^{\infty } \left| c_{n} \right|^{2}$ is normalizable.
The two questions has some nuanced differences. We do not know
whether another situation could happen. For example, its energy is
infinity, while its wave function is still normalizable. This
situation may happen under some conditions, but in this paper we
assume, to QKR, the energy finity is equivalent to the wave function
normalizability. And the situation of energy finity and wave
function normalizability as time runs to the infinity is referred as
dynamic localization.

There are close analogies between QKR and some problems in the solid
state physics. The quasienergy band of the Floquet operator with
rational $\frac{\tau}{2\pi}$ is analogous to the energy band of the
Bloch operator. The irrational case Floquet operator has been
transformed into an Anderson localization problem in
\cite{Fishman1982}. But there is one fundamental difference. An
Bloch electron can be in the eigenstate of the Bloch operator, while
QKR can never be in a Floquet eigenstate, because a Bloch eigenstate
has a finite energy, while in the case of QKR the a Floquet
eigenstate has an infinite energy. So the rotator state will always
be a superposition of many Floquet eigenstates. The absolute values
of the coefficients of the eigenstates in the superposition never
change, and what changes is only the relative phases between
different Floquet eigenstates.

\subsection{\label{sec:level2}Inverse Cayley transform and dynamic localization}

One way to study QKR is to find all the eigenvalues and eigenstates
of the Floquet operator. If the eigenstates are extended, dynamic
delocalization happens and \textit{vice versa}. Milek \textit{et al}
has formally proved absolutely continuous spectra of the Floquet
operator imply delocalization. When the Floquet operator has
singular continuous spectra, the usual dynamic delocalization does
not happen \cite{Milek1989}. Can the Floquet operator have extended
eigenstates and singular continuous spectra under some conditions,
for example, when $\frac{\tau}{2\pi}$ is a Liouville number?
Jitomirskaya \textit{et al} proved almost Mathieu equation has
singular continuous spectra under some conditions
\cite{Jitomirskaya1994}. Yet we assume localized eigenstate is
equivalent to the dynamic localization to QKR.

One way to find all the eigenstates of the Floquet operator is to
transform the Floquet eigenstate problem into time independent
Hermitian operator eigenstate problem. Especially when
$\frac{\tau}{2\pi}$ is irrational it is transformed into an Anderson
problem \cite{Fishman1982}. The Floquet eigenstate problem is
\begin{equation}
F \phi_{\lambda}=e^{-i V(\theta)} e^{-i H_{0} \tau} \phi_{\lambda} =e^{-i \lambda} \phi_{\lambda}.
\end{equation}

A Hermitian operator $U$ can be transformed into a unitary operator
$O$ by Cayley transform $O=\frac{1+iU}{1-iU}$. A unitary operator
$O$ can be transformed into a Hermitian operator by inverse Cayley
transform $U=i\frac{1-O}{1+O}$. In complex analysis, Cayley
transform, which is a linear fractional transform, maps the real
line to the unit circle on the complex plane and the inverse Cayley
transform maps the unit circle to the real line. Cayley transform
and its inverse transform of both the operator form and the complex
number form has many applications in different fields of
mathematics.

Substitute
\begin{equation}
e^{-i V(\theta)}=e^{-ik\cos\theta}=\frac{1+i U(\theta)}{1-i
U(\theta)},
\end{equation}
where $U(\theta)=-\tan(\frac{V(\theta)}{2})=-\tan(\frac{k}{2}\cos
\theta)$, into Eq. $(12)$. We get
\begin{equation}
T_{m} u_{\lambda,m}+\sum_{r} U_{r} u_{\lambda,m+r} = 0,
\end{equation}
where $u_{\lambda}=(1+e^{i (\lambda-H_{0})}) \phi_{\lambda}$, $U_{r}
= U_{-r}$ is the Fourier coefficient of $U(\theta)$, and $T_{m}=\tan
\frac {\lambda-\tau m^{2}} {2}$. QKR is transformed into a particle
moving on a periodic or non-periodic lattice depending on $\tau$.
The evolution of QKR has no direct relation with the new particle on
a lattice. But the eigenstates of the Floquet operator and the
eigenstates of the new formed Hermitian operator are related by
$u_{\lambda}=(1+e^{i (\lambda-H_{0})}) \phi_{\lambda}$. If
$\phi_{\lambda}$ is extended or localized, then $u_{\lambda}$ is
also extended or localized and \textit{vice versa}.

We define A as
\begin{equation}
\left(
\begin{array}{lllllll}
 \cdots & \cdots & \cdots & \cdots & \cdots & \cdots & \cdots \\
 \cdots & T_{-2} & U_1 & U_2 & U_3 & \cdots & \cdots \\
 \cdots & U_{-1} & T_{-1} & U_1 & U_2 & U_3 & \cdots \\
 \cdots & U_{-2} & U_{-1} & T_0 & U_1 & U_2 & \cdots \\
 \cdots & U_{-3} & U_{-2} & U_{-1} & T_1 & U_1 & \cdots \\
 \cdots & \cdots & U_{-3} & U_{-2} & U_{-1} & T_2 & \cdots \\
 \cdots & \cdots & \cdots & \cdots & \cdots & \cdots & \cdots
\end{array}
\right).
\end{equation}
Then
\begin{equation}
A u_{\lambda}+U_0 u_{\lambda}=0
\end{equation}

One basic observation is the diagonal matrix elements $T_{m}$ of $A$
are pseudorandom numbers if $\frac{\tau}{2\pi}$ is irrational.
$T_{m}$ is a Cauchy distribution. Because $U(\theta)$ looks quite
regular here, now we invoke Anderson's result \cite{Anderson1958}.
$u_{\lambda}$ is localized. The above analysis gives an analytical
proof of dynamic localization discovered in \cite{Casati1979}.
Fishman, Grempel, and Prange's inspiring result connects the fields
of quantum chaos and Anderson localization in disordered matters. It
points out the physical mechanism of dynamic localization of QKR.
Inverse Cayley transform seems to be the only method to transform
the Floquet operator eigenstate problem to the disordered
Hamiltonian eigenstate problem.

\subsection{\label{sec:level2}Flaws in the inverse Cayley transform method}

Now we check the above argument step by step. The equidistribution
of the sequence $\{ \alpha n^{2} \emph{Mod} 1\}_{n}$ when $\alpha$
is irrational can be proved from the equidistribution of sequence
$\{ \alpha n \}_{n}$ and van der Corput's Theorem. While the
equidistribution of sequence $\{ \alpha n \emph{Mod} 1 \}_{n}$ can
be proved by Weyl criterion. So all the sequences $\{ \alpha n^{m}
\emph{Mod} 1 \}_{n}$, where is $m$ is a positive integer and
$\alpha$ irrational, are equidistributed between 0 and 1. Other
equidistribution examples include $\{ x^{n} \emph{Mod} 1 \}_{n}$,
where $x>1$. The sequence of all multiples of $\alpha$ by all prime
numbers $\{ 2 \alpha, 3 \alpha, 5 \alpha, 7 \alpha, 11 \alpha,
\ldots \emph{Mod} 1 \}$ is studied in \cite{TaoMa2007QKPR} as the
quantum kicked prime number rotator (QKPR). It localizes when
$\alpha$ is irrational because $\{ 2 \alpha, 3 \alpha, 5 \alpha, 7
\alpha, 11 \alpha, \ldots \emph{Mod} 1 \}$ are equidistributed
between $0$ and $1$. QKPR also localizes when $\alpha$ is rational
such as $\frac{1}{3}$ because now it is analogous to a generalized
kicked dimer model.

The number theoretic property of a number sequence is used in Shor's
algorithm to find prime factors of a composite number $S$. If $S$ is
a large number, and we define a rotator with energy levels $\{ p^{n}
Mod S \}_{n}$, where $n$ runs from 1 to $S$ and $1<p<S$. If $S$ is a
prime number, dynamic localization (To a finite Hilbert space, this
is not very rigorous) also happens. If $S$ is a composite number,
dynamic localization generally does not happen. Although we can not
find the exact prime factors of $S$, at least we can judge whether
$S$ is prime.

There is one inherent paradox (Local pseudorandomness paradox) of
localization caused by number theoretic randomness. On the one hand,
the randomness of the sequence $\{\frac{\tau}{4\pi}n^2 \}_n$ with
irrational $\frac{\tau}{4\pi}$, is only meaningful when $n$ goes to
infinity. On the other hand, if the wave is localized, how could the
wave feel the randomness of $\frac{\tau}{4\pi} n^2$ of very large
$n$, where there is randomness? This can be said in another
language, how can a remote (global) randomness influence a localized
wave or how can a localized wave feel a remote (global) randomness?
In Anderson localization, every random variable is independent. So
in the place where the wave is localized, the randomness is real
randomness. But in number theoretic randomness, there are strong
correlations between the local pseudorandomness. For example, the
QKR wave function is localized at several localization lengths. Is
the pseudorandomness at several localization lengths randonmness or
even pseudorandomness? Let's take $k=1$. So the wave function is
localized between several basis vectors. For example, they are from
$|-10\rangle$ to $|10\rangle$. Is the sequence
$\{\frac{\tau}{4\pi}n^2 \}_n$ from $n=-10$ to $10$ pseudorandomness?

Randomness can be classified into two categories, the statistical
randomness and pseudorandomness caused by chaos or number theoretic
origin. The local pseudorandomness paradox is not a problem which
only concerns QKR or Eq. $(16)$. It is also not a new question. L.
Boltzmann tried to found statistical mechanics on the ergodic
theory. A recent example is Bohigas-Giannoni-Schmit conjecture
\cite{BGS1984}.

The first problem of the inverse Cayley transform method to prove
localization is to some irrational $\frac{\tau}{4\pi}$, the sequence
$\{ \frac{\tau}{4\pi} n^{2} \}_{n}$ may be not random enough to
cause localization. This is the pseudorandomness problem that has
been discussed by many authors, such as \cite{Fishman1982,
Casati1984, Casati1998}. In \cite{TaoMa2007IUMM}, we give an example
of irrational $\frac{\tau}{4\pi}$ which is not irrational enough to
cause localization.

Second, Eq. $(16)$ is not really an Anderson localization problem.
In Anderson localization, we know the Hamiltonian, and we do not
know the eigenvalues and eigenstates. While in Eq. $(16)$, we know
the eigenvalue beforehand, which is just $-U_0$. While the unknown
$\lambda$ is contained in the diagonal matrix elements $T_{m}$. This
is not an Anderson localization problem. It is rather an inverse
Anderson localization problem. We know one special eigenvalue and we
need to find what diagonal ``disorders'' satisfies Eq. $(16)$.
Surely there are infinitely many diagonal ``disorders'' satisfies
Eq. $(16)$. Eq. $(16)$ has the form of an eigenvalue problem, while
in fact it is a nonlinear equation of $\lambda$ or $e^{-i \lambda}$.
One may say to a general irrational $\frac{\tau}{2\pi}$, whatever
$\lambda$ is, the sequence $T_{m}$ is random enough to guarantee the
eigenstate localized. In Anderson's formulation of localization, the
diagonal matrix elements are required to be independent random
variables \cite{Anderson1958}. No independent random variables can
satisfy an equation, in which these independent random variables are
dependent variables. We use the terms independent and dependent
variables in their original sense of \textit{Statistics}.

Generally an infinite matrix has infinite eigenvalues and
eigenstates. For example, we find a $\lambda$ which satisfies Eq.
$(17)$. Then the operator $A$ defined in $(16)$ surely has many
other eigenvalues besides $-U_0$. Are other eigenstates of the
infinite matrix with definite $\lambda$ also localized?

The third problem is $U(\theta)=-\tan(\frac {k} {2}\cos \theta)$ is
regular only if $k<\pi$. When $k \geq \pi$, $U(\theta)$ is
discontinuous at some points. The first discontinuous point is
$\theta=\arccos \frac {\pi} {k}$. At this point $U(\theta)$ is
$\infty$. In \cite{Anderson1958}, the interaction is required to be
``falling off as the distance $r\rightarrow \infty$ faster than
$1/r^{3}$''. In \cite{Anderson1958}, $r$ is the distance between two
lattice sites, while $r$ in $U_{r}$ is the index. But both two
express the interaction strength between different sites. Does
$U_{r}$ fall faster than $1/r^{3}$? This question can be answered in
two ways. First, if $U_{r}\leq \frac {c} {r^{3}}$, where $c$ is a
constant, then $U(\theta)=\sum_{r} U_{r} e^{i r \theta}$ is
convergent at every point of  $\theta$. So $U_{r}$ can not fall
faster than $1/r^{3}$. Second, when $k \geq \pi$, $U(\theta)$ is not
square integrable. We only need to check whether $U(\theta)^2$ is
integrable in the small domain $\arccos \frac {\pi} {k} -\delta <
\theta< \arccos \frac {\pi} {k} + \delta$, where $\delta$ is a small
number. If $k > \pi$, in the domain, $\frac {k} {2}V(\theta)$ can be
approximated by $\frac{\pi}{2}+c \theta $, where $c$ is the
derivative of $\frac {k} {2}V(\theta)$ at the point $\theta=\arccos
\frac {\pi} {k}$. But $\tan (\frac{\pi}{2}+c \theta )$ is not square
integrable in the domain. If $k=\pi$, $\tan(\frac {1} {2} k
V(\theta))^{2}$ is approximated by $(\tan(\frac {\pi} {2}
(1-\frac{\theta^{2}}{2})))^{2}=(\cot(\frac{\pi}{4}x^{2}))^{2}
\approx 1/(\sin \frac{\pi x^{2}}{4})^{2} \approx 1/( \frac{\pi
x^{2}}{4})^{2}$. It is not square integrable either. So when $k \geq
\pi$, $U(\theta)$ is not square integrable. $\sum_{r}U_{r}^{2}$ is
divergent due to Parseval's identity. The worst estimate is $U_{r}$
falls slower than $1/\sqrt{r}$. The Fourier expansion $U_r$ of
$U(\theta)$ does not exist at all, when $k=\pi\times Odd$, where
$Odd$ is an odd positive integer such as $1,3,5,\ldots$.

So when $k \geq \pi$, we can not invoke Anderson's result to prove
the vectors $u_{\lambda}$ are localized in Eq. $(16)$. This is a
fatal flaw of the inverse Cayley transform method to prove dynamic
localization. In one dimension it is easy to localize
\cite{Lee1985}. So the slowly falling $U_r$ may not destroy
localization. But in the conventional theory of condensed matters,
the strongest long range force is Coulomb force $\propto
\frac{1}{r}$. QKR with a general irrational number
$\frac{\tau}{2\pi}$ seems to always localize however large $k$ is.
We think the real meaning of the breakdown of the inverse Cayley
transform method is there may be a localization-delocalization
transition when $k$ increases from $k<\pi$ to $k>\pi$ and $k=\pi$ is
the critical point to many kicked systems. In the QKPR of
\cite{TaoMa2007QKPR}, there is apparently a
localization-delocalization transition in the rational case
$\tau=\frac{2\pi}{3}$. When $k=1$ the prime number rotator
localizes, while when $k=5$ it delocalizes.

There are several flaws of the inverse Cayley transform method. Why
not develop an independent localization theory concerning Floquet
operators or unitary operators? And the new theory which treats time
dependent problems (unitary operators) is parallel with the Anderson
localization theory which treats time independent problems
(Hermitian operators). This does not seem to be an easy task. The
first difficulty is there is not a unitary perturbation theory
concerning unitary operators, while to Hermitian operators there are
many perturbation methods. We will return to the problem of
developing a dynamic localization theory of the Floquet operator in
the second paper.

\section{\label{sec:level1}Anomalous localization}
The inverse Cayley transform method provides a physical picture,
because of diagonal pseudorandomness, the eigenstate tends to be
localized. But there is another path of the development of QKR
theory, which gives a different picture. Casati and Guarneri
proposed there is a non-empty set of irrational $\frac{\tau}{2\pi}$,
to which dynamic localization can not happen \cite{Casati1984}. We
must resolve these two conflicting views. Fishman \textit{et al}
also pointed out to the Liouville number $\frac{\tau}{2\pi}$, things
are very delicate and they excluded the Liouville number
$\frac{\tau}{2\pi}$ in their theory \cite{Fishman1982}.

We have to discriminate two kinds of different ``delocalizations''.
To general irrational $\frac{\tau}{2\pi}$, the localization length
$l_{\tau}$ is estimated to be $\alpha D$ \cite{Chirikov1981,
Chirikov1988}, where $D$ is the classical diffusion constant
$\frac{k^{2}}{2}$. And the factor $\alpha$ is estimated to be
$\frac{1}{2}$ in \cite{Shepelyansky1986, Fishman1989}. The first
kind of ``delocalization'' is $\alpha D<l_{\tau}<\infty$, which is
actually localization with an anomalous localization length. This is
referred as anomalous localization. The second delocalization is
$l_{\tau}=\infty$. The anomalous localization can be seen from
examples such as $\frac{\tau}{4\pi}=\frac{1}{3}+\frac{\sqrt{5}-1}{2
\times 10^{m}}$ or the continued fraction $\frac{\tau}{4\pi}=
(0;3,10^m,1,1,1,\cdots)$, where m is a large number in both cases.

But we do not know all the irrational numbers $\frac{\tau}{2\pi}$
with anomalous localization or delocalization. We do not know to a
special $\frac{\tau}{2\pi}$ whether anomalous localization or
delocalization depends on $k$. For example, when $k<\pi$,
delocalization does not happen to a $\tau$, while when $k>\pi$ it
does. We do not know whether both cases can happen to general
Liouville numbers, such as Liouville constant
$L=\sum_{j=1}^{\infty}10^{-j!}$. The continued fractional of L is
$\{ 0; 9, 11, 99, 1, 10, 9, 999999999999, \ldots \}$. The $n$-th
incrementally largest term consisting only of 9s occurs precisely at
position $2^{n}-1$, and this term consists of $(n-1)n!$ 9s
\cite{Shallit1982}.

Although we have constructed an irrational $\frac{\tau}{2\pi}$ with
delocalization in \cite{TaoMa2007IUMM}. There is not one specific
irrational $\frac{\tau}{2\pi}$ that has been numerically calculated
to be delocalized. It is difficult to do a numerical simulation to
these special irrational numbers. It is the periodic structure of
the Floquet operator which causes delocalization. So a successful
calculation must preserve the periodic structure. To the rational
$\frac{\tau}{2\pi}=\frac{p}{q}$ case, the truncated Hilbert space is
at least as large as several $q$s. This makes it very difficult to
calculate the rational cases with large denominators.

\subsection{\label{sec:level2}A quantum analogy of Lyapunov exponent equation}

There are two different Floquet operators $F$ and $F'$. We calculate
the difference between $U(N)=F^{N}$ and $U(N)'=F'^{N}$. We define
$\delta U(N)=U(N)'-U(N)$, $\delta F=F'-F$ and $\delta
\tau=\tau'-\tau$. \begin{equation}
\begin{split}
&\delta U(N)= F' U(N-1)'-F U(N-1) \\
&=F' U(N-1)'-F' U(N-1)  \\
&\quad +F' U(N-1)-F U(N-1) \\
&=F' (U(N-1)'-U(N-1))+\delta F U(N-1) \\
&=F' \delta U(N-1)+\delta F U(N-1) \\
&=\sum_{j=1}^{N} U(N-j)' \delta F U(j-1). \\
\end{split}
\end{equation}
In the last step, the mathematical induction method is used. Eq.
$(17)$ describes the divergence between unitary operators because of
small difference $\delta F$. This is a quantum analogy of the
Lyapunov exponent equation. The classical Lyapunov exponent equation
describes the divergence between two orbits in the phase space
because of small difference between the initial conditions. The
stability of quantum systems is indicated by their sensitivity to
the perturbation of the Hamiltonian \cite{Peres1984}. We think the
sensitivity of the Floquet operator to a perturbation is another
criteria to judge whether chaos is relevant to a quantum system. In
other words, the stability of a periodically driven quantum system
is decided by the sensitivity of the Floquet operator to
perturbation. There are two kinds of sensitive perturbation to $F$.
The first kind is the phase sensitivity. The perturbation rotates
the phase of some matrix elements by an angle $\pi$. The second kind
is the perturbation greatly modifies the module of some matrix
elements. In QKR, $\delta F$ is not sensitive to $\delta k$ or
$\delta \hbar$. It is sensitive to $\delta \tau$. We consider
$\delta \tau=\tau'-\tau$. The matrix element
\begin{equation}
\begin{split}
\delta U(N)_{nm} &= \sum_{j=1}^{N} \sum_{l} (U(N-j)' e^{-iV})_{nl}\\
&\times(e^{-1/2i l^{2} \tau'}-e^{-1/2i l^{2} \tau}) U(j-1)_{lm}.
\end{split}
\end{equation}
For simplicity, we now assume $n=0$. The summation of $l$ of
$U(j-1)_{lm}$ is effective in a finite bandwidth, which is at most
from $-jk$ to $jk$. $|e^{-1/2i l^{2} \tau'}-e^{-1/2i l^{2} \tau}|
\approx 1/2l^{2} \delta \tau$. The row $(U(N-j)' e^{-iV})_{nl}$ and
the column $U(j-1)_{lm}$ are unit vectors. From Cauchy-Schwarz
inequality,
\begin{equation}
\begin{split}
&|\sum_{l} (U(N-j)' e^{-iV})_{nl} (e^{-1/2i l^{2} \tau'}-e^{-1/2i
l^{2} \tau}) U(j-1)_{lm} |^2\\
&\leq (\sum_{l}| (U(N-j)'e^{-iV})_{nl}|^2)\\
&\times(\sum_{l}|(e^{-1/2i
l^{2} \tau'}-e^{-1/2i l^{2} \tau}) U(j-1)_{lm}|^2) \\
&= \sum_{l}|(e^{-1/2i l^{2} \tau'}-e^{-1/2i l^{2} \tau})
U(j-1)_{lm}|^2 \\
&\leq \sum_{l}|1/2l^2\delta\tau U(j-1)_{lm}|^2\\
&=\sum_{l}1/4 l^4 \delta\tau^2 | U(j-1)_{lm}|^2 \\
&\leq 1/4 (j
k)^{4} \delta \tau^2,
\end{split}
\end{equation}
where the equal sign $=$ in the last step comes from when
$|U(j-1)_{lm}|^2$ concentrates on the boundary $jk$ of the
bandwidth, which is actually impossible.
\begin{equation}
\begin{split}
&|\sum_{l} (U(N-j)' e^{-iV})_{nl} (e^{-1/2i l^{2} \tau'}-e^{-1/2i
l^{2} \tau}) \\
&U(j-1)_{lm} | \leq 1/2 (j k)^{2} \delta \tau.
\end{split}
\end{equation}
After the summation of $j$ in Eq. $(18)$,
\begin{equation}
|\delta U(N)_{nm}| \leq 1/6 N^{3}k^{2} \delta \tau.
\end{equation}
Before the time $(\frac{6 \epsilon}{k^{2} \delta \tau})^{1/3}$,
$|\delta U(N)_{nm}| \leq \epsilon$. Eq. $(21)$ is first got by
Casati \textit{et al} \cite{Casati1984}. In fact, generally
\begin{equation}
|\delta U(N)_{nm}| \ll 1/3N^{3}k^{2} \delta \tau.
\end{equation}
To the general rational number $\frac{\tau}{2\pi}$, the diffusion
speed is far slower than the case $\tau \emph{Mod}4\pi=0$. So the
effective bandwidth of $U(j-1)$ is far smaller than $jk$. The
diffusion speed $v$ of the case $\tau=0$ is $k$. To a rational
$\frac{\tau}{2\pi}$, the diffusion speed is $v_{\tau}$. Then the
bandwidth of $U(j-1)$ is $v_{\tau}j$. So
\begin{equation}
|\delta U(N)_{nm}| \leq 1/6 N^3 v_{\tau}^{2} \delta \tau.
\end{equation}
When $t/\tau < (\frac{6 \epsilon}{v_{\tau}^2 \delta \tau})^{1/3}$,
$|\delta U(N)_{nm}| \leq \epsilon$. We can define $(\frac{6
\epsilon}{v_{\tau}^2 \delta \tau})^{1/3} \tau$ as the divergence
time, after which the $U(N)$ and $U(N)'$ are significantly different
from each other. $v_{\tau}$ is quite small except some strongly
resonant cases. To the resonant case $\frac{\tau}{2\pi}=\frac{1}{q}$
with $k=1$, we estimate $v_{\tau}$ can be as small as
$c_{1}e^{-c_{2}q}$, where $c_1$ and $c_2$ are two constants, the
relation between which and $q$ and $k$ is unclear at present
\cite{TaoMa2007IUMM}.

In the derivation of Eq. $(21)$, the summation of $l$ is from $-jk$
to $jk$. This is the upper limit of diffusion with the fastest
diffusion speed, which happens only when $\tau=0$ or $4\pi$. The
more exact value of the bandwidth is $\frac{e}{2}jk$ from the
property of Bessel function, where $e=2.71828\cdots$. So a factor
$\frac{e^2}{4}=1.84726\cdots$ should be added to Eq. $(21)$ and
$(23)$. We do not consider the factor, which will cause qualitative
differences to our discussion.

The above calculation crucially depends on the almost band structure
of $F$, $F'$, $U(N)$ and $U(N)'$. A band matrix can not be a unitary
matrix. The different columns or rows of a unitary matrix are
orthonormal with each other. A band matrix can never has its columns
or rows orthonormal with each other. For example, let $b$ denote the
end of one row of a band matrix. Then another row of the band matrix
has its start at $b$. These these two rows can not be orthonormal
with each other. Another thing worth notice is $e^{-iV}$ is a
Toeplitz matrix. $U$ in Eq. $(14)$ is also Toeplitz. Toeplitz
matrices are quite rigid. The quantum kicked harmonic oscillator
\cite{Berman1991} does not have such a nice property. Casati
\textit{et al} tried to explain dynamic localization from the
perspective of band random matrix \cite{Casati1990}. The Floquet
operator of QKR is neither band nor random. When $k>\pi$, from the
perspective of Anderson localization, $u_\lambda$ should be
delocalized, but it is still localized. The mechanism of the
localization of Eq. $(16)$ and QKR is even stronger than the
mechanism of Anderson localization. It is this mechanism that QKR
theorists need to find.

We can get a more exact estimate than Eq. $(21)$ under some
conditions. In Eq. $(19)$, we use Cauchy-Schwarz inequality to drop
$U(N-j)'$. We can also drop $U(j-1)$. A better estimation is to drop
the unitary matrix with a larger bandwidth. For example, if
$\frac{\tau}{2\pi}$ and $\frac{\tau'}{2\pi}$ are all irrational
numbers, the summation of $l$ is restricted to the localization
length $l_{\tau}$.
\begin{equation}
|\delta U(N)_{nm}| \leq 1/6 N l_{\tau}^2 \delta \tau.
\end{equation}
So the divergence time between different irrational
$\frac{\tau}{2\pi}$ are far larger than the divergence time between
rational and irrational $\frac{\tau}{2\pi}$.

Eq. $(24)$ also holds when a rational $\frac{\tau}{2\pi}$ is close
to a general irrational number. For example
$|\frac{\tau}{2\pi}-\frac{\sqrt{5}-1}{2}|<\epsilon$, where
$\frac{\tau}{2\pi}$ is rational and $\epsilon$ small.

\subsection{\label{sec:level2}Analytical and numerical proof of anomalous localization}
Now we estimate the anomalous localization length. From Eq. $(23)$,
before the divergence time
\begin{equation}
t=(\frac{6 \epsilon}{v_{\tau}^2 \delta \tau})^{1/3} \tau,
\end{equation}
QKRs of $\tau$ and $\tau'$ will not diverge from each other much. We
assume $\frac{\tau}{2\pi}$ is rational and $\frac{\tau'}{2\pi}$
irrational. So QKR of $\tau'$ will at least diffuse to the length
$(\frac{6 \epsilon v_{\tau}}{\delta \tau})^{1/3}$. We assume
$\epsilon=1$. Then the localization length $l_{\tau'}$ is at least
$(\frac{6 v_{\tau}}{\delta \tau})^{1/3}$. Another method is from Eq.
$(24)$, if the divergence time is $N\tau$, then $N =
\frac{6}{l_{\tau'}^2 \delta\tau}$. And $l_{\tau'}=Nv_{\tau}$. We get
the same result. So
\begin{equation}
l_{\tau'}=(\frac{6 v_{\tau}}{\delta \tau})^{1/3}.
\end{equation}
We assume $v_{\tau}(k)=v_{\tau}(k=1)k$. Then
\begin{equation}
l_{\tau'}=(\frac{6 v_{\tau}(k=1) k}{\delta \tau})^{1/3}.
\end{equation}

The anomalous localization is significant when $v_{\tau}$ is large,
and $\delta\tau$ small. $v_{\tau}$ is large when $\tau$ is close to
strong resonances and $k$ large. In our numerical calculation, we
take $\frac{\tau'}{4\pi}$ is the sum of $\frac{1}{3}$ and a small
number. $\tau=4\pi\frac{1}{3}$ has the fastest delocalization speed
except $\tau=0$ or $4\pi$. In our calculations, $\hbar=1$, $k=1$ and
the truncated Hilbert space is from $|-500\rangle$ to $|500\rangle$.

In FIG. 1 and 2, $\tau=4\pi\times [0;3,100,1,1,1,\cdots]$, where the
continued fraction $[0;3,100,1,1,1,\cdots]\approx 1/3-0.00110064$.
The localization length is around $6$ from FIG. 1 and 2.

In FIG. 3 and 4, $\tau=4\pi\times [0;3,1000,1,1,1,\cdots]$, where
$[0;3,1000,1,1,1,\cdots]\approx 1/3-0.00011101$. The localization
length is around $15$ from FIG. 3 and 4. Notice there is staircase
like structure in FIG. 3. It is very rough. We do not know what they
are.

In FIG. 5 and 6, $\tau=4\pi\times [0;3,10000,1,1,1,\cdots]$, where
$[0;3,10000,1,1,1,\cdots]\approx 1/3-0.00001111$. The localization
length is around $36$ from FIG. 5 and 6. Note far outside the
localization length, the wave function is not smooth.

\begin{figure*}
\begin{center}
   \begin{minipage}{16.2 cm}
   \includegraphics[width=5.3 cm]{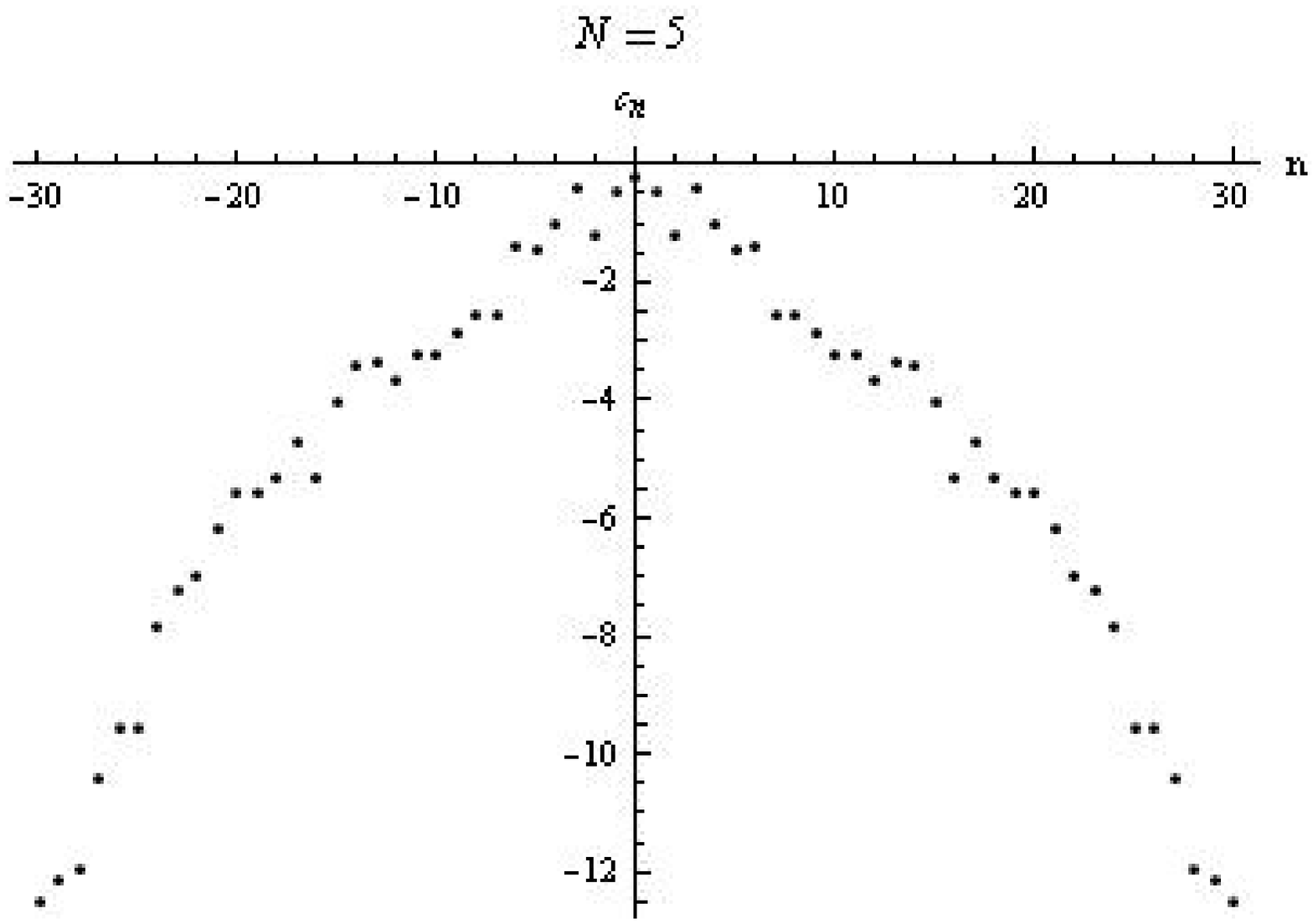}
   \includegraphics[width=5.3 cm]{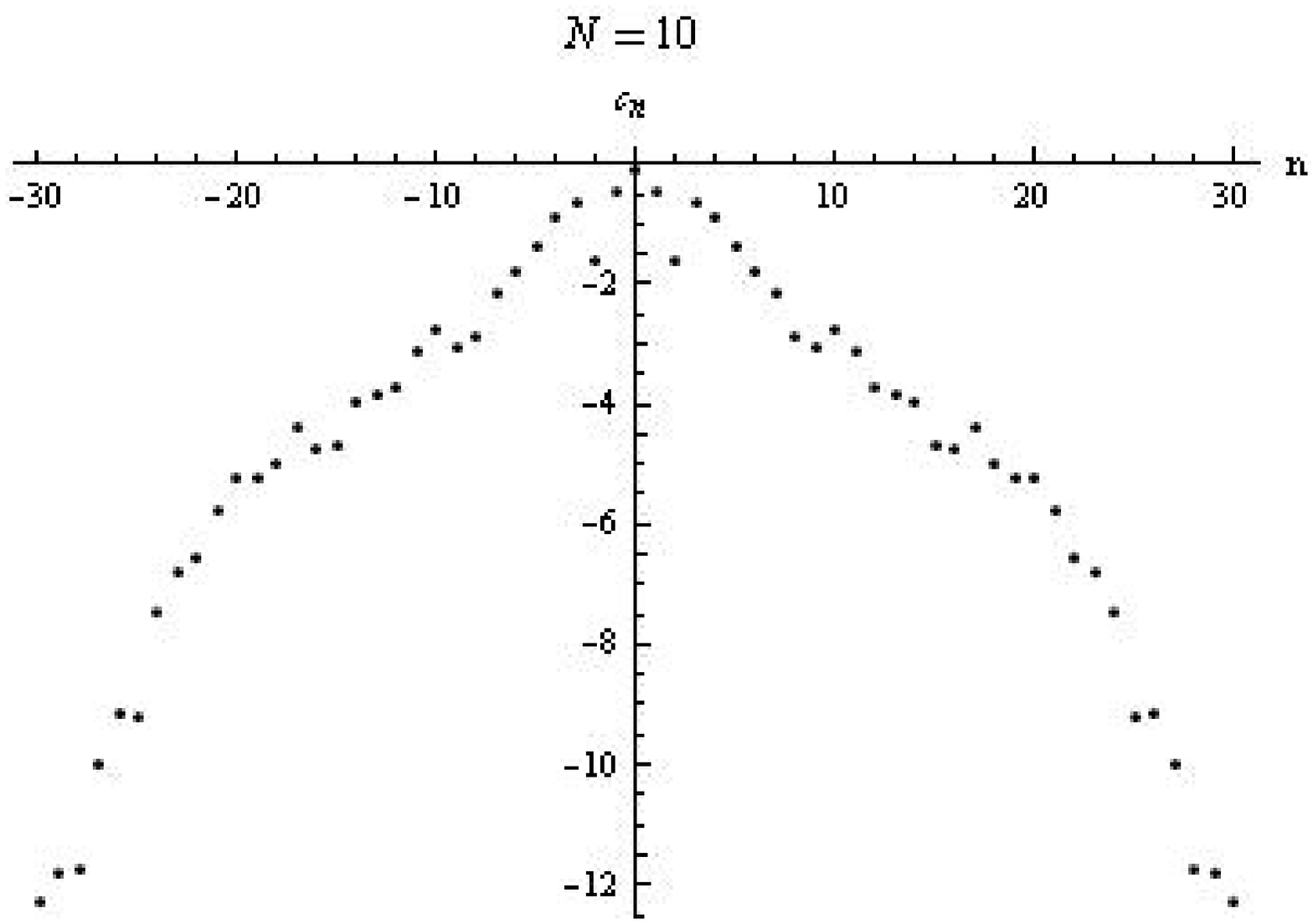}
   \includegraphics[width=5.3 cm]{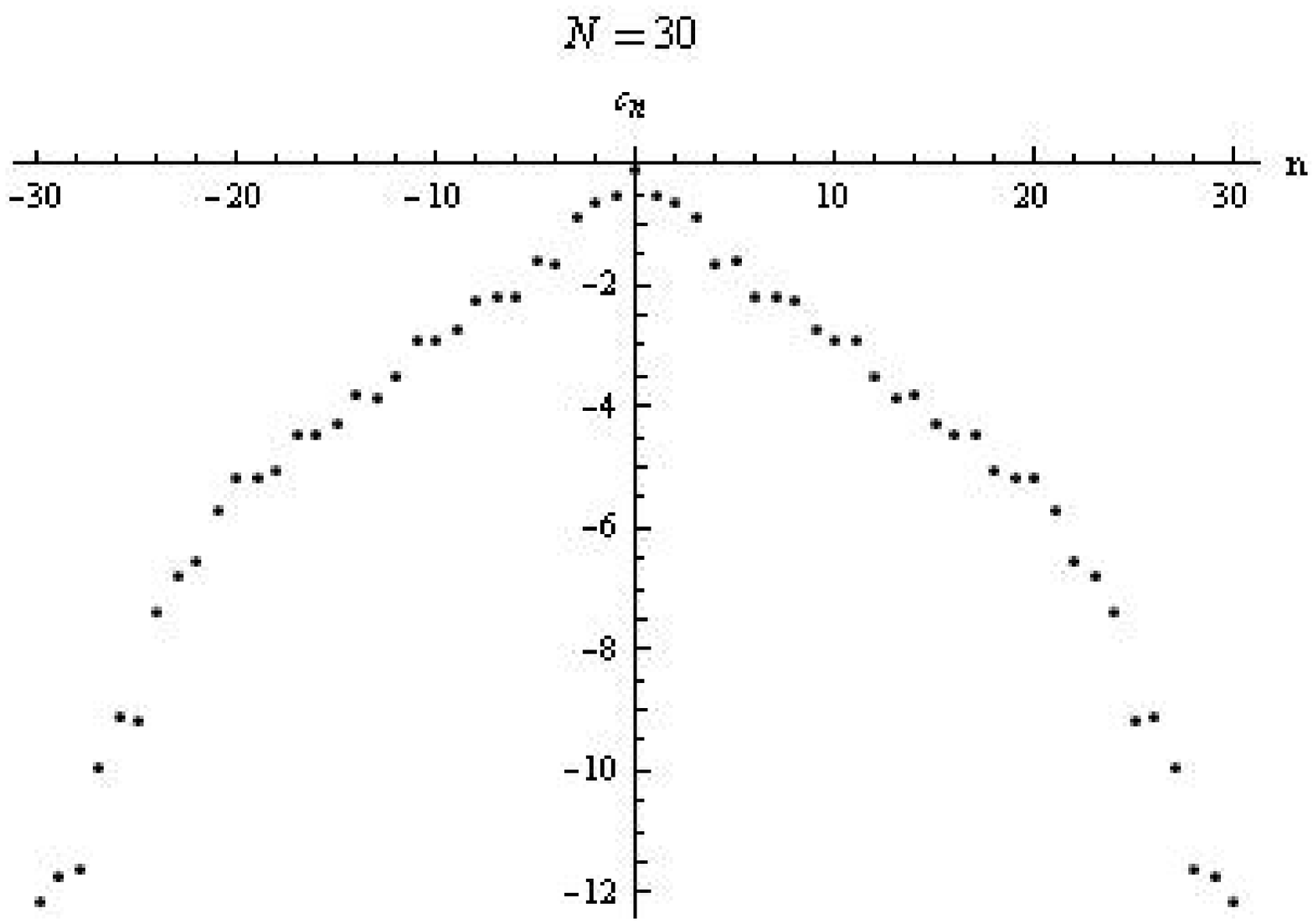}
\caption{QKR wave function at different time. $N$ indicates at time
$2^N\tau$. For example, $N=10$ means at time $2^{10}\tau$. $n$ is
the $n$-th basis $|n \rangle = \frac{1}{\sqrt{2 \pi }} e^{in\theta}$
and $c_n$ is the base-10 logarithm of the absolute value of the wave
function on the $n$-th basis. $k=1$, $\tau=4\pi\times
[0;3,100,1,1,1,\cdots]$. The initial state is $|0\rangle$.}
\end{minipage}
   \begin{minipage}{16.2 cm}
   \includegraphics[width=5.3 cm]{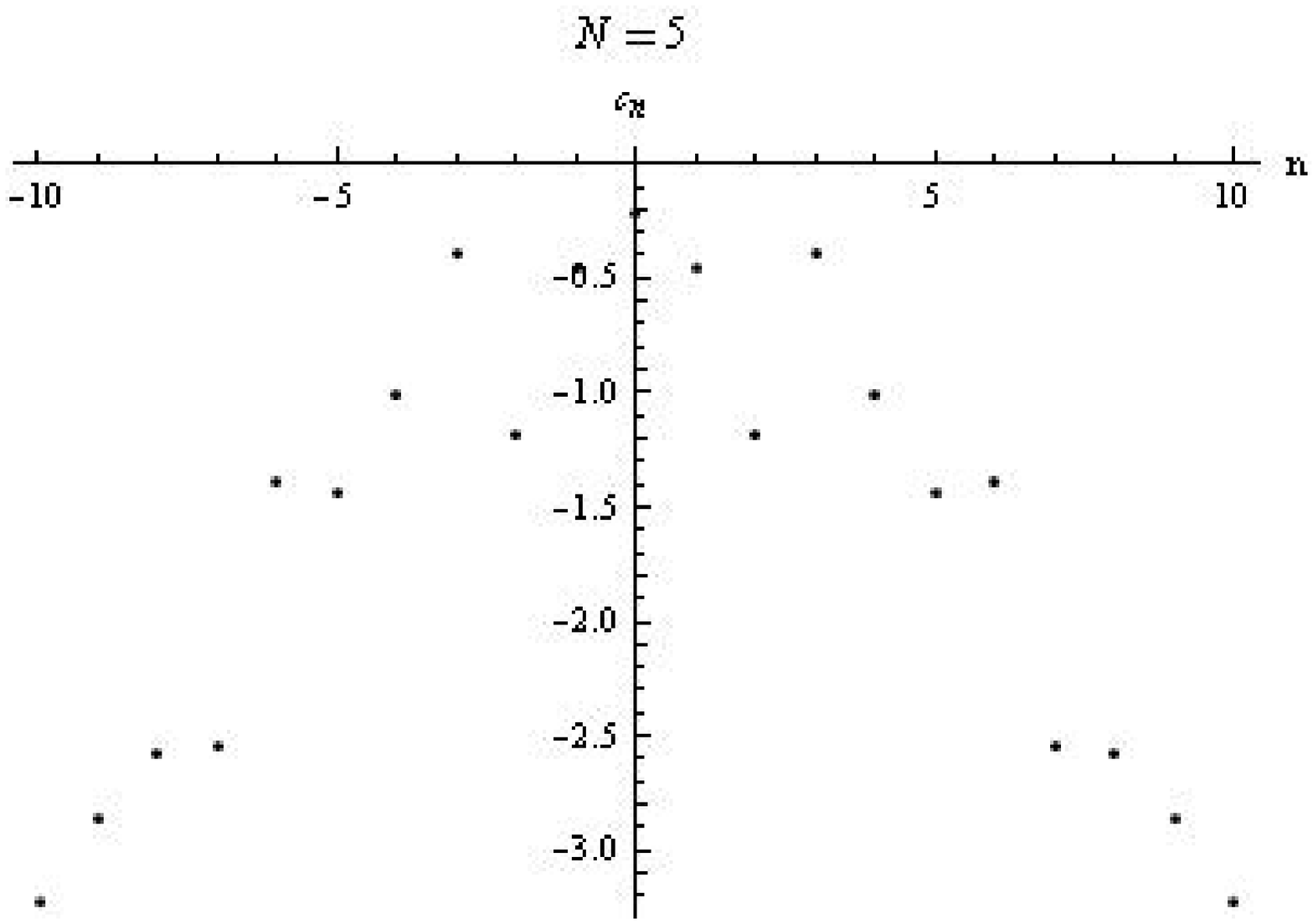}
   \includegraphics[width=5.3 cm]{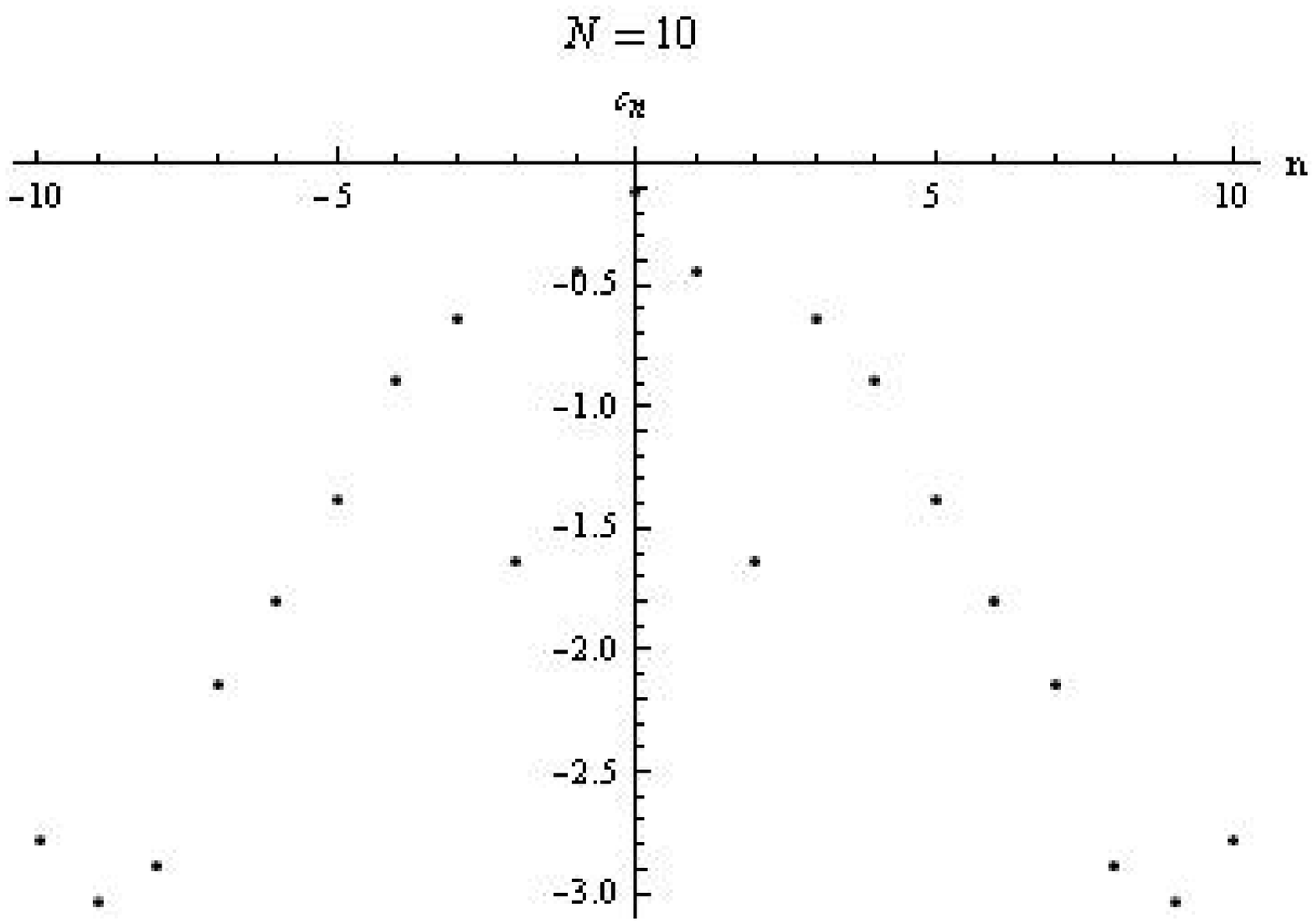}
   \includegraphics[width=5.3 cm]{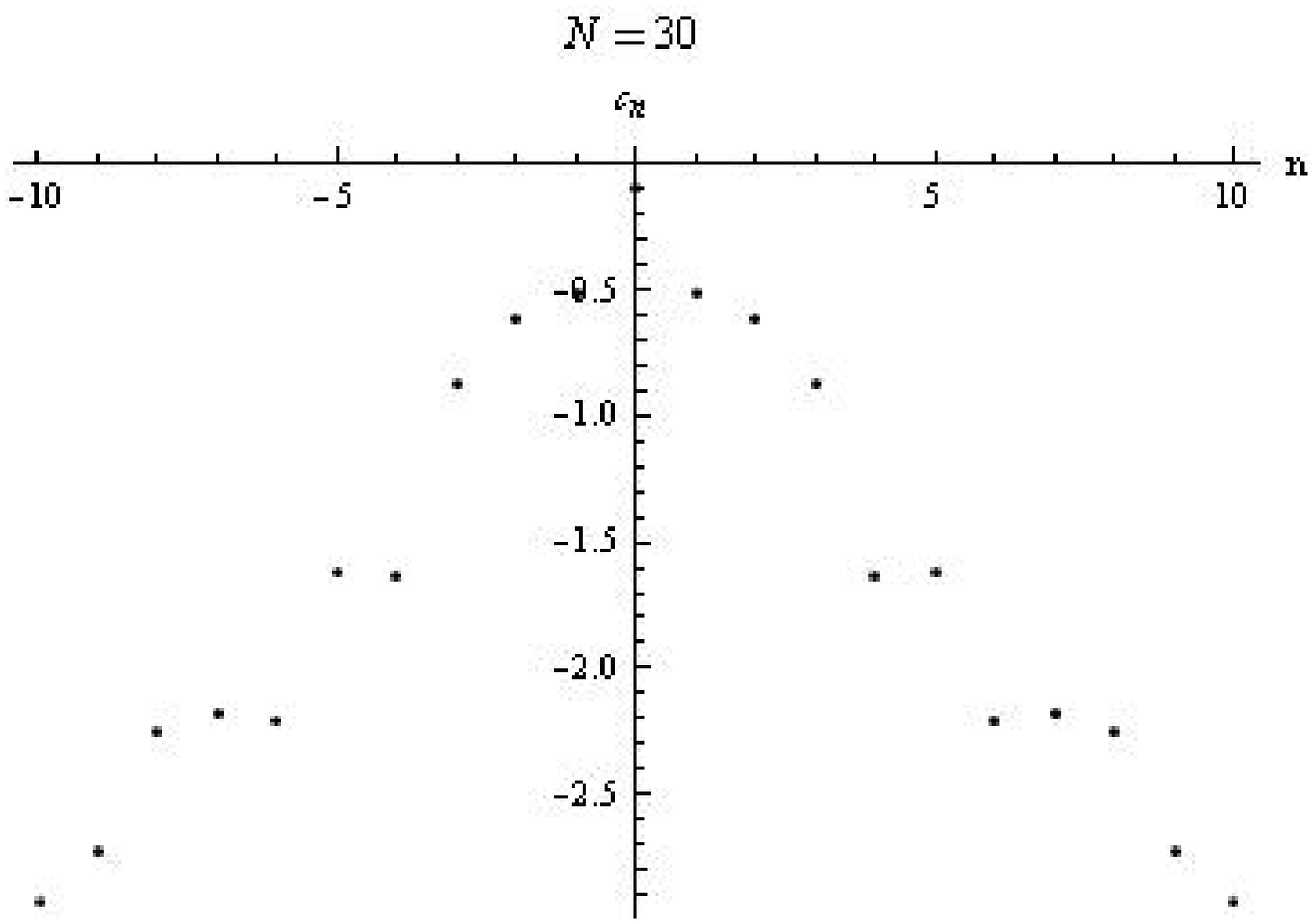}
\caption{QKR wave function at different time. FIG. 1 and FIG. 2 are
the same calculation.}
\end{minipage}
\end{center}
\end{figure*}

\begin{figure*}
\begin{center}
   \begin{minipage}{16.2 cm}
   \includegraphics[width=5.3 cm]{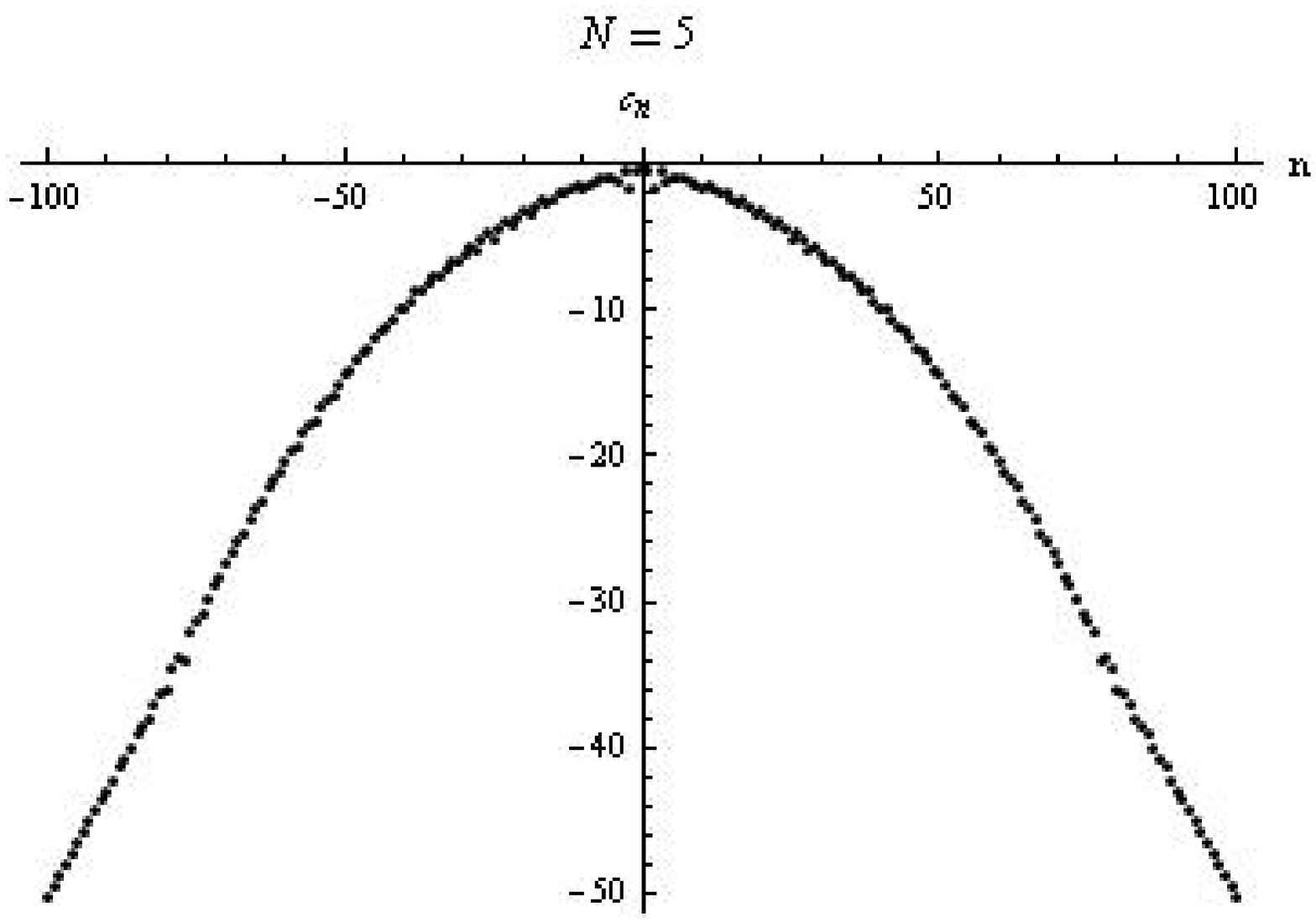}
   \includegraphics[width=5.3 cm]{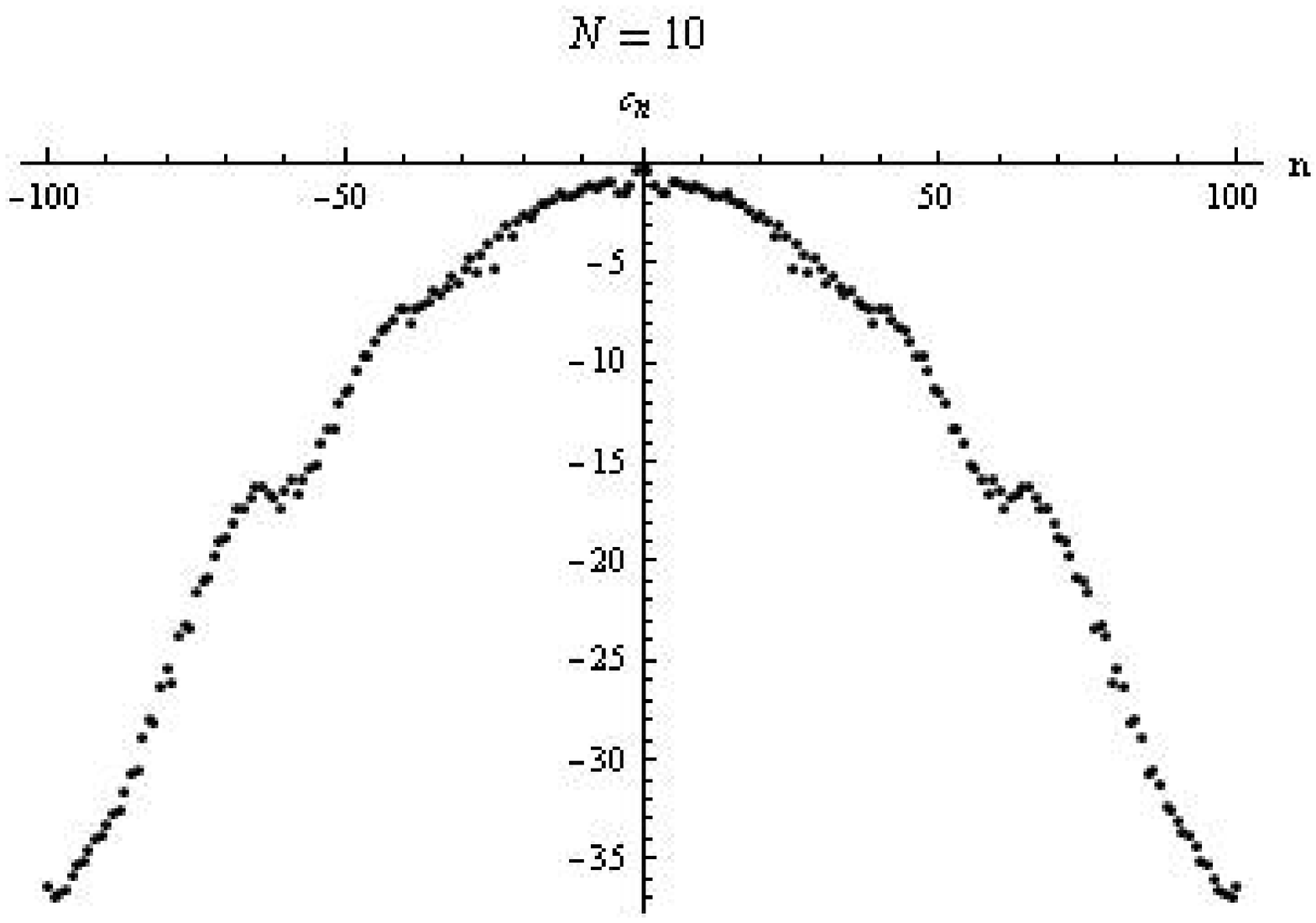}
   \includegraphics[width=5.3 cm]{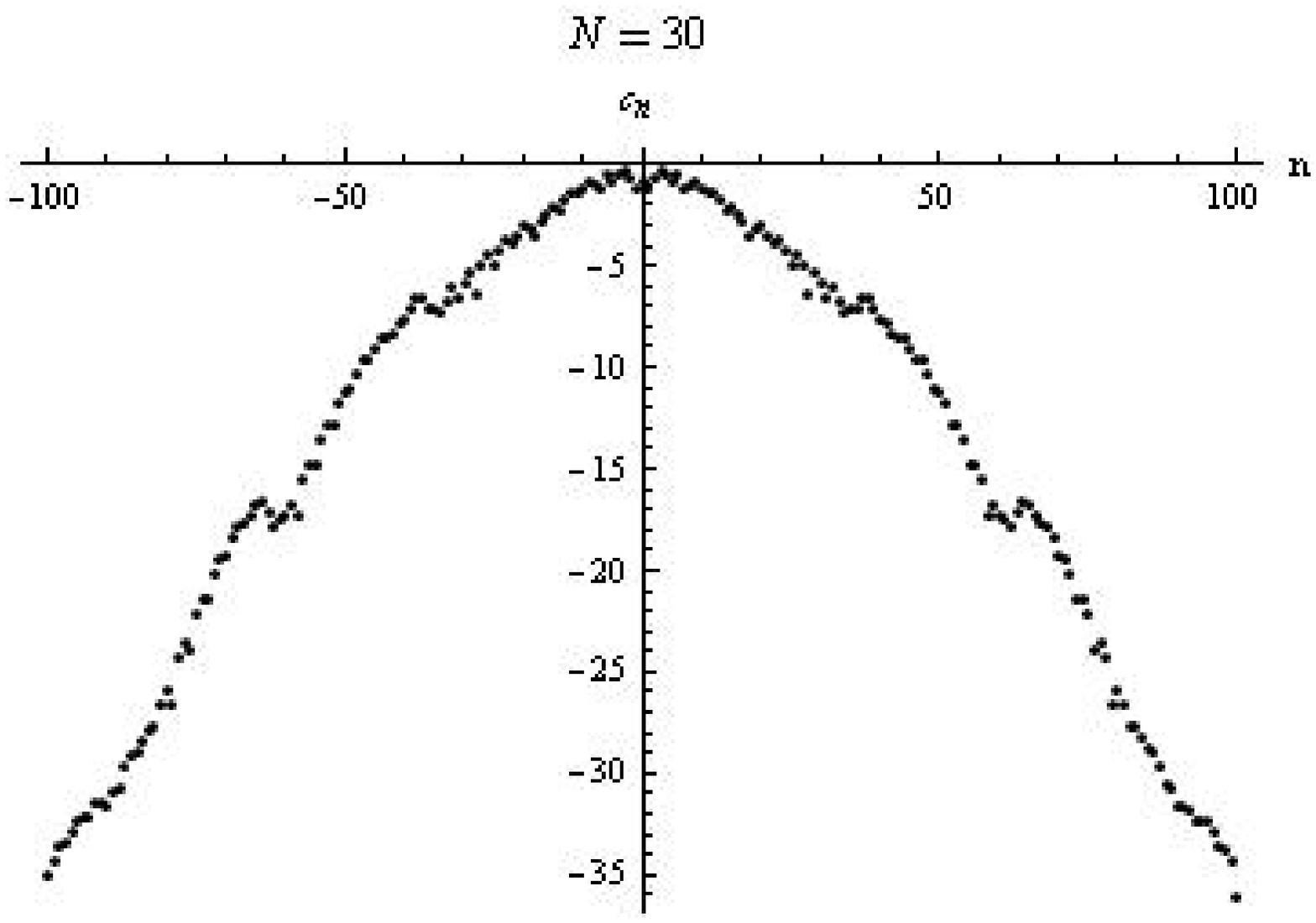}
\caption{QKR wave function at different time. $k=1$,
$\tau=4\pi\times [0;3,1000,1,1,1,\cdots]$. The initial state is
$|0\rangle$. }
\end{minipage}
   \begin{minipage}{16.2 cm}
   \includegraphics[width=5.3 cm]{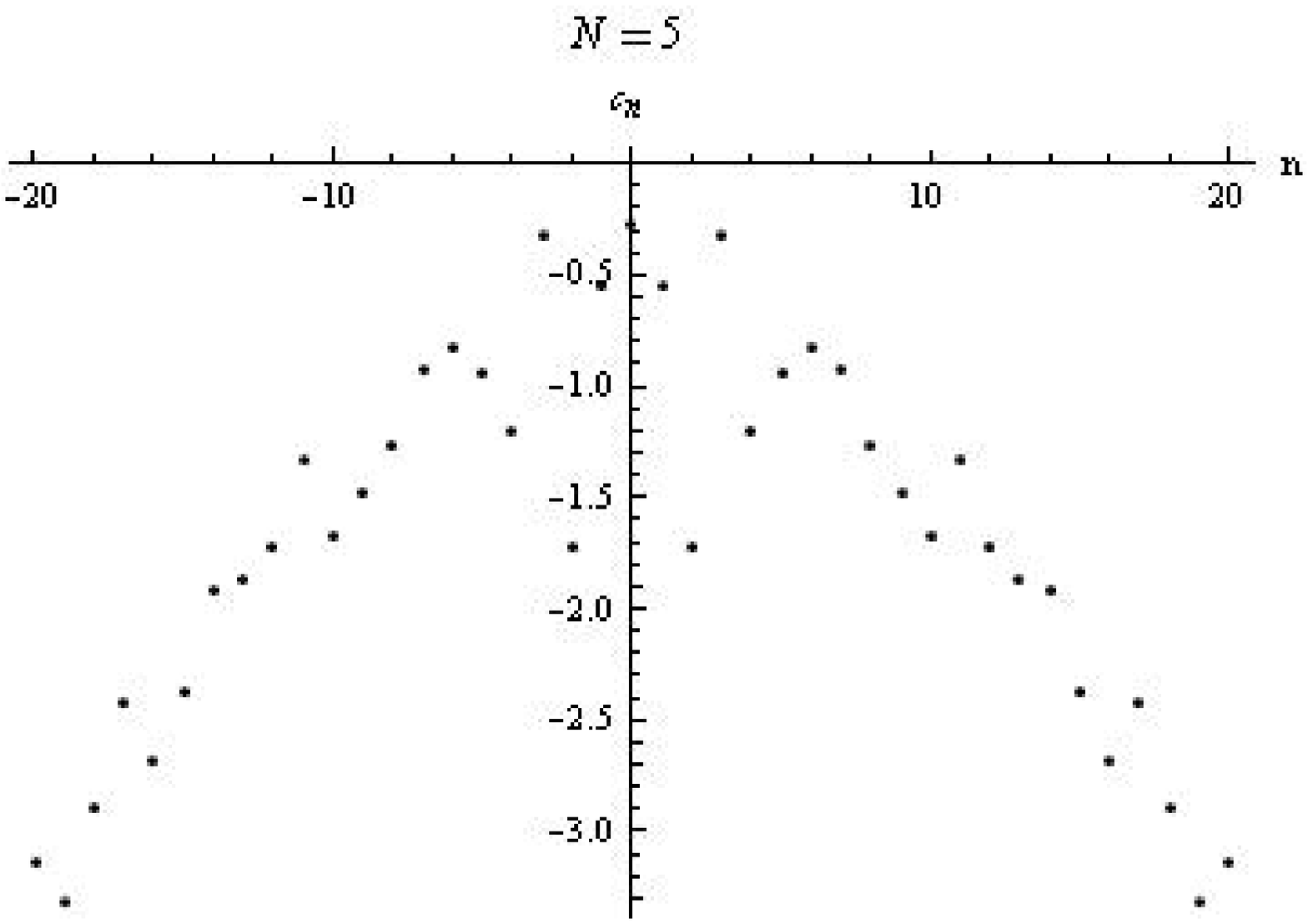}
   \includegraphics[width=5.3 cm]{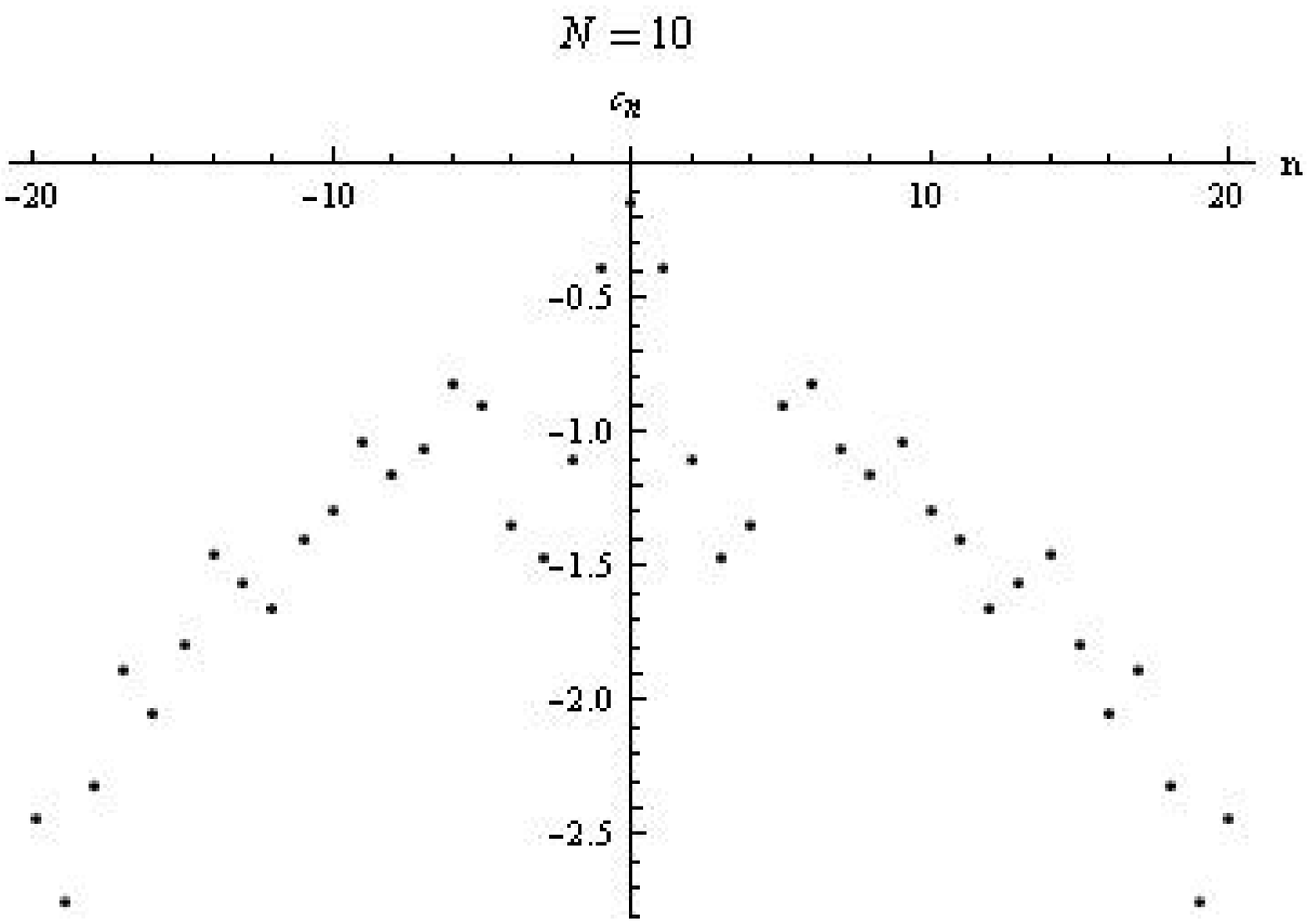}
   \includegraphics[width=5.3 cm]{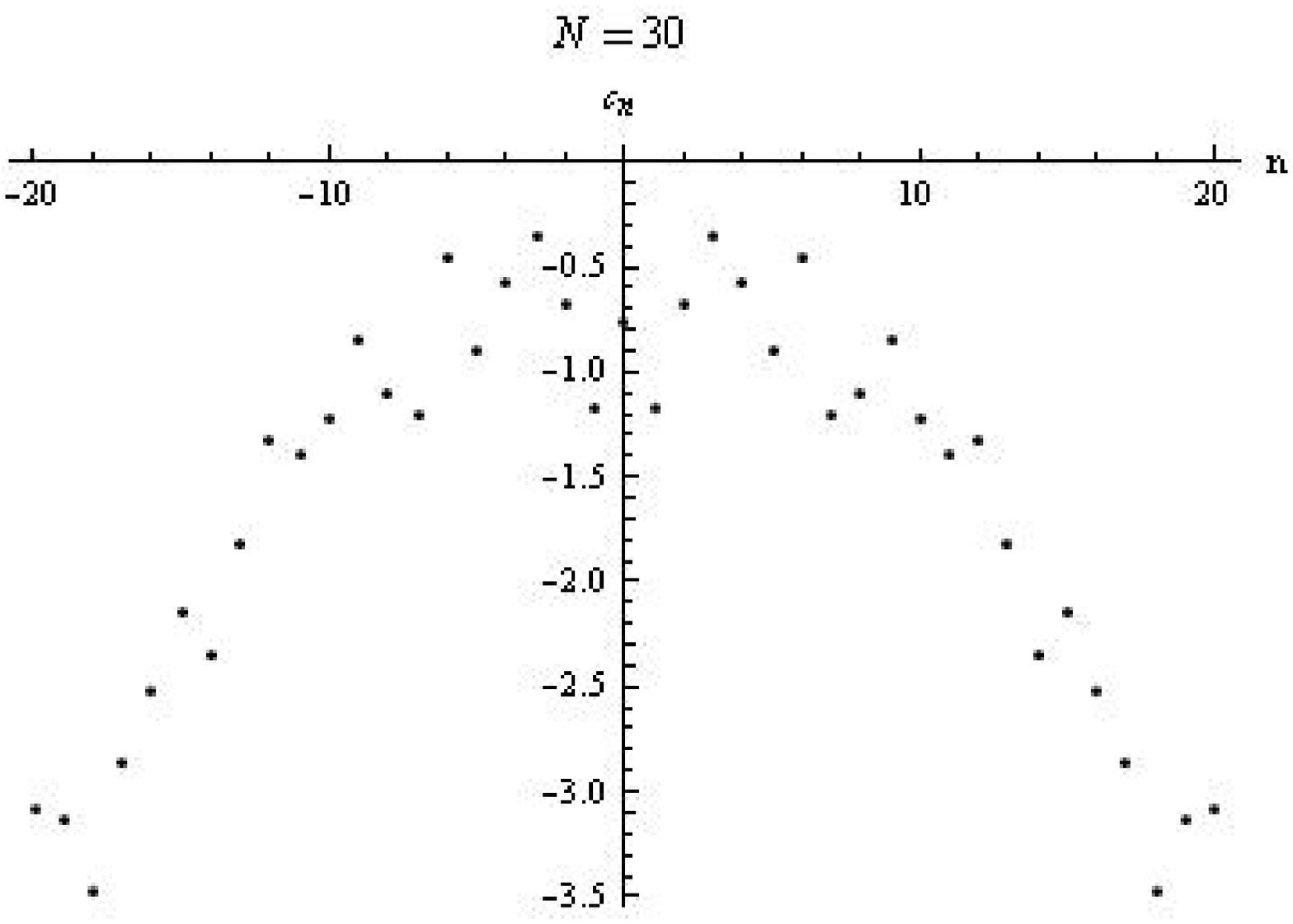}
\caption{QKR wave function at different time.  FIG. 3 and FIG. 4 are
the same calculation.}
\end{minipage}
\end{center}
\end{figure*}

\begin{figure*}
\begin{center}
   \begin{minipage}{16.2 cm}
   \includegraphics[width=5.3 cm]{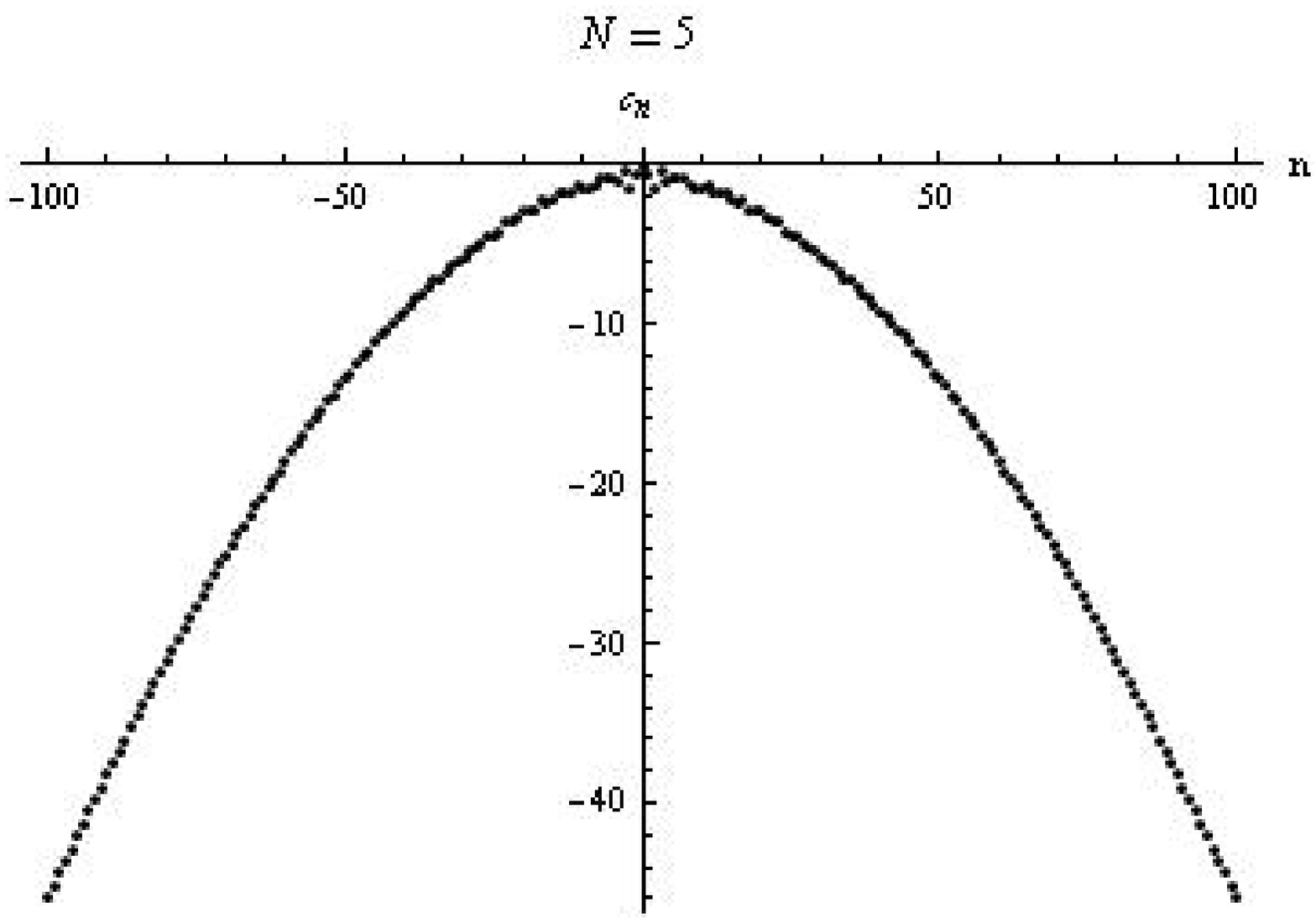}
   \includegraphics[width=5.3 cm]{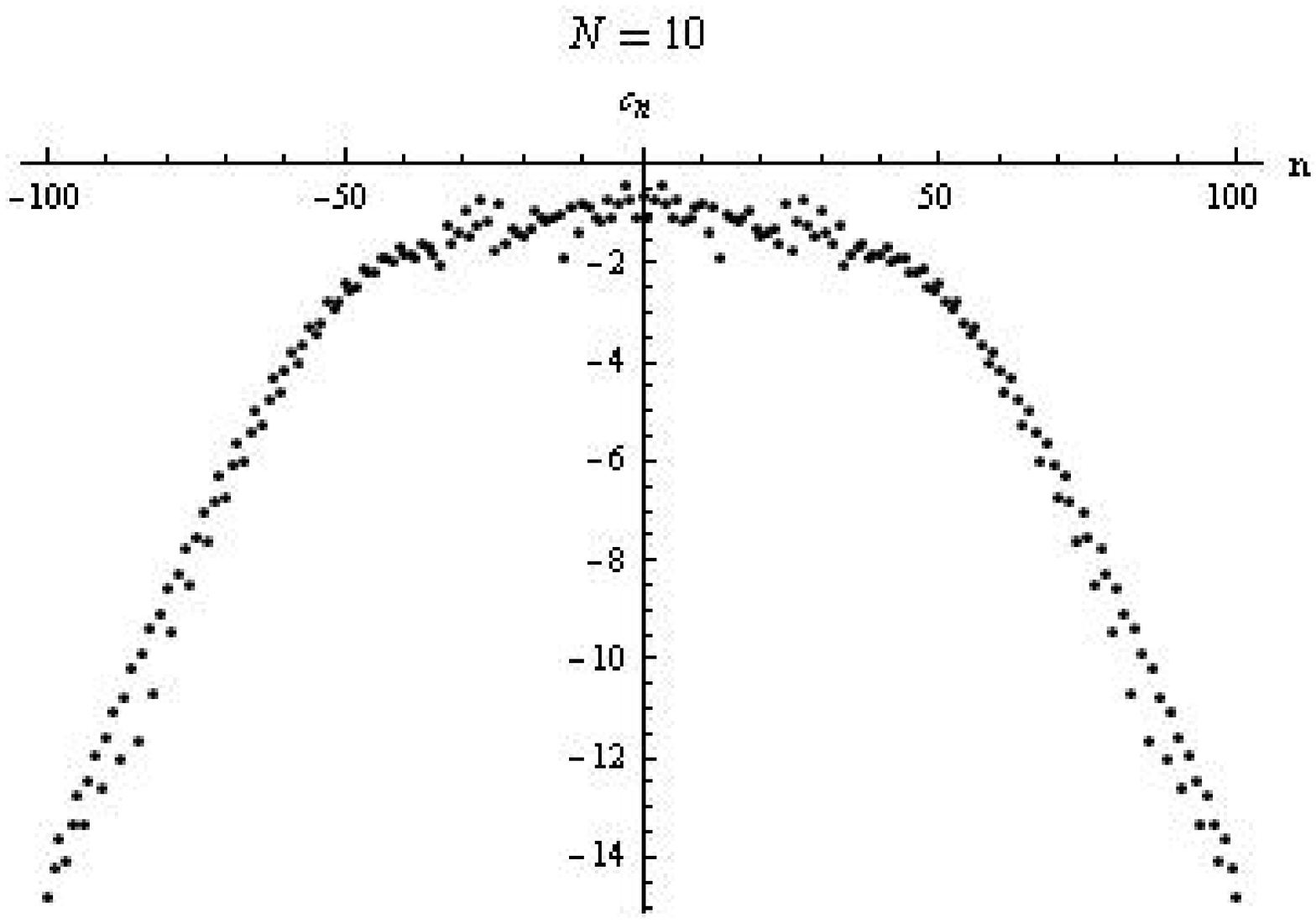}
   \includegraphics[width=5.3 cm]{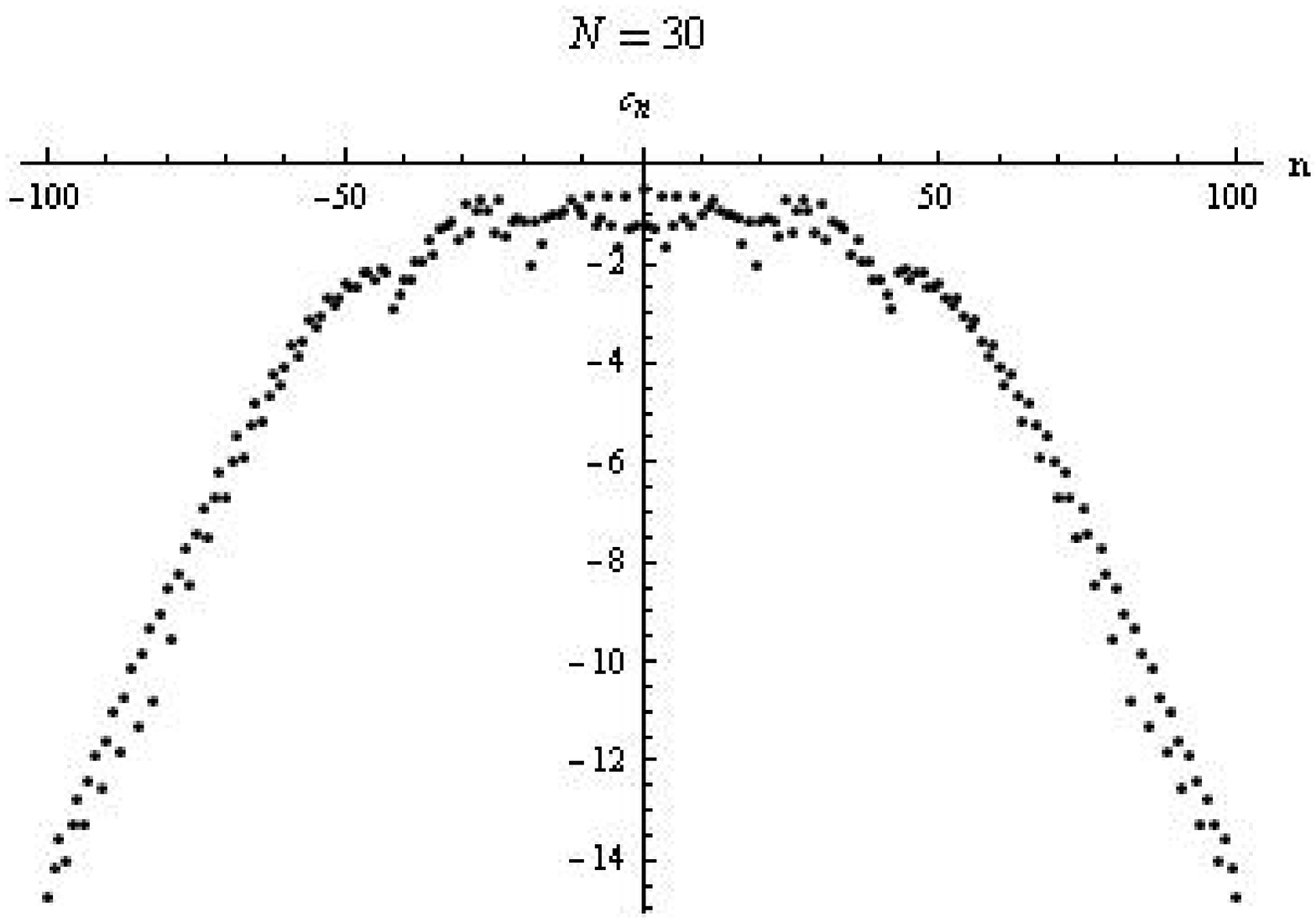}
\caption{QKR wave function at different time. $k=1$,
$\tau=4\pi\times [0;3,10000,1,1,1,\cdots]$. The initial state is
$|0\rangle$.}
    \end{minipage}
    \begin{minipage}{16.2 cm}
   \includegraphics[width=5.3 cm]{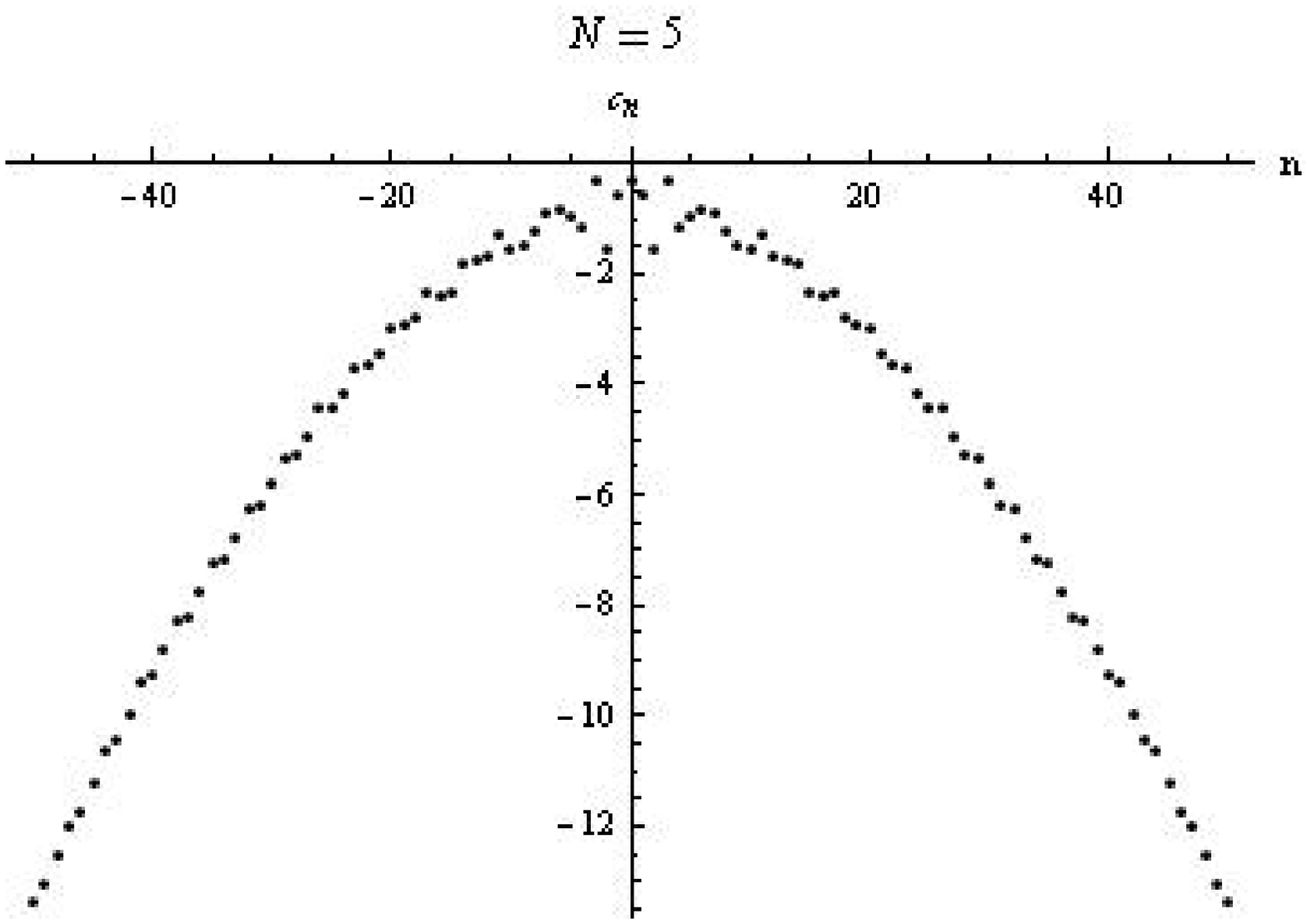}
   \includegraphics[width=5.3 cm]{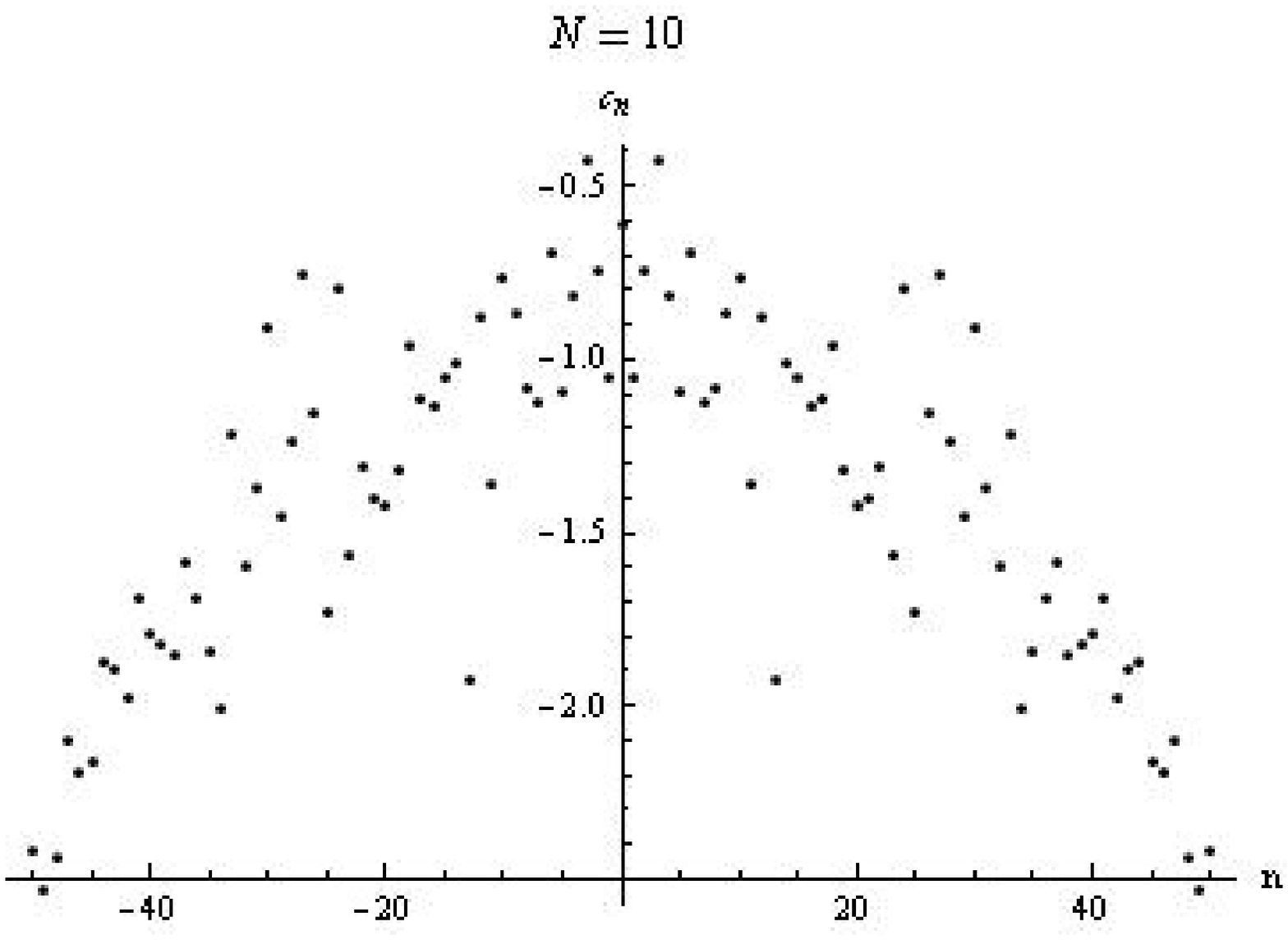}
   \includegraphics[width=5.3 cm]{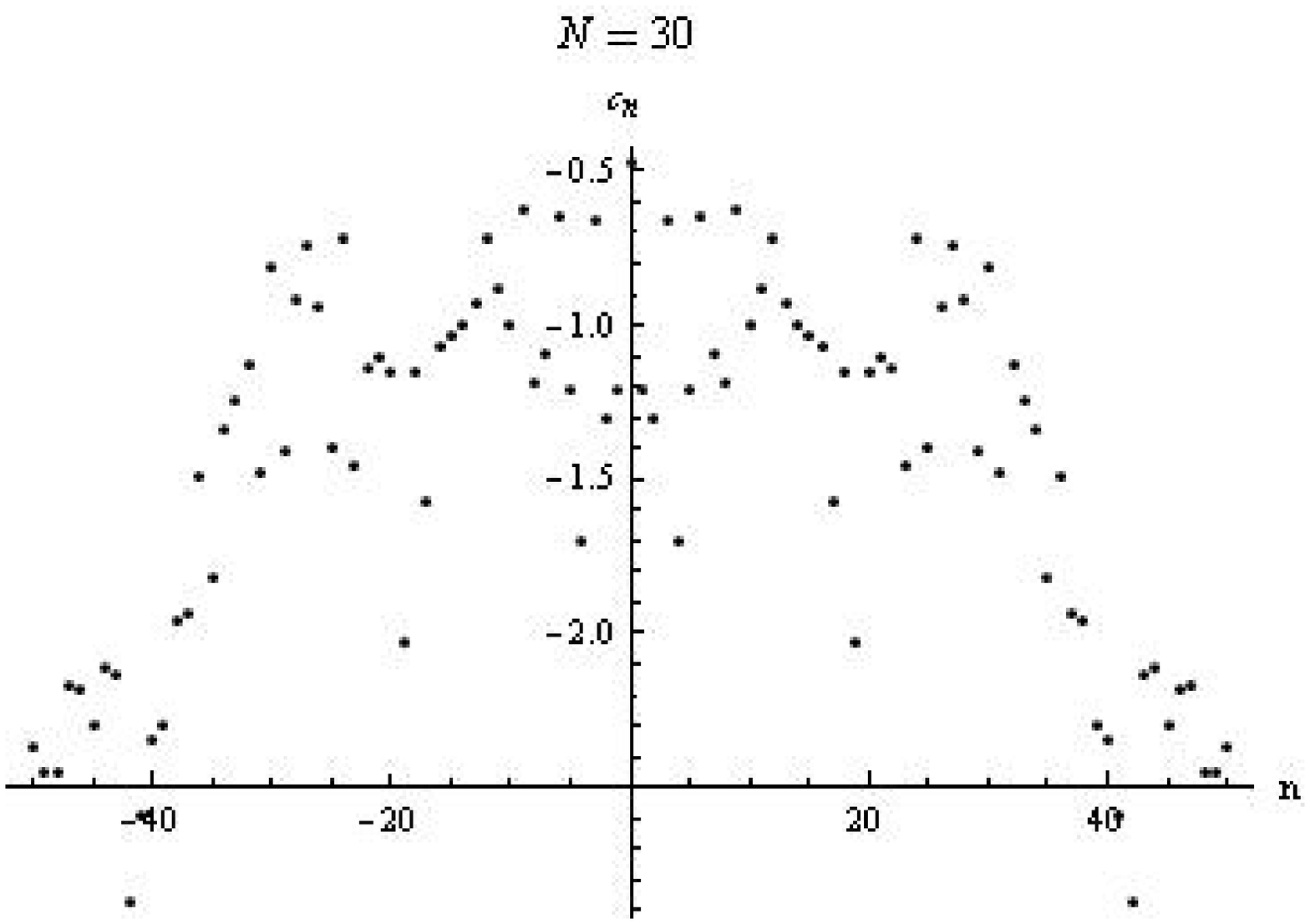}
\caption{QKR wave function at different time.  FIG.5 and FIG.6 are
the same calculation.}
\end{minipage}
\end{center}
\end{figure*}

It is difficult to tell the exact localization length, which seems
to change at different times. Compare FIG. 2, 4, and 6; the general
trend of increasing localization length is clear. Compare these
localization lengths together. $6^3:15^3:36^3=1:15.6:216$. The
general trend $l_{\tau'} \propto \delta\tau^{-1/3}$ is confirmed. It
seems $\frac{l_{\tau'}}{\delta\tau^{1/3}}$ is a monotonically
increasing function of $\delta\tau^{-1/3}$.

\subsection{\label{sec:level2}Casati and Guarneri's argument}
In \cite{Casati1984}, Casati and Guarneri defined a quantity
$R_{\tau}(n,\psi)=\frac{1}{n}\sum_{j=1}^{n}|(F_{\tau}^{n}
\psi,\psi)|^{2}$ to measure the recurrent behavior of QKR. They
proposed if the Floquet operator $F_{\tau}$, where $\tau$ is the
kick period, has purely continuous spectrum, then when $n
\rightarrow \infty$, $R_{\tau}(n,\psi) \rightarrow 0$. It follows
from Eq. $(21)$ that $|R_{\tau'}(n,\psi)-R_{\tau}(n,\psi)| \leq 2
\gamma n^{3}$. Then starts Casati and Guarneri's argument.
\begin{equation}
\begin{split}
R_{\tau'}(N_{n},\psi)
&= |R_{\tau'}(N_{n},\psi)-R_{\tau_{n}}(N_{n},\psi)|+R_{\tau_{n}}(N_{n},\psi)\\
&= \gamma |\tau'-\tau_{n}| N_{n}+R_{\tau_{n}}(N_{n},\psi)
\end{split}
\end{equation}
Find the sequences $\tau_{n}$ and $N_{n}$. They satisfy the
conditions, when $n \rightarrow \infty$, $|\tau'-\tau_{n}| N_{n}
\rightarrow 0$ (Condition 1) and $R_{\tau_{n}}(N_{n},\psi)
\rightarrow 0$ (Condition 2) at the same time. Then
$R_{\tau'}(N_{n},\psi) \rightarrow 0$. In this way they argue there
are some irrational $\frac{\tau'}{2\pi}$s, which have a purely
continuous measure.

For a long time, Casati and Guarneri's surprising argument puzzles
QKR theorists \cite{Casati1998}. One gap in the argument is whether
when $n \rightarrow \infty$, Condition $1$ and $2$ are true at the
same time. $\tau_{n}$ is not a static rational number, so
$R_{\tau_{n}}(N_{n},\psi) \rightarrow 0$ is true only if $N_{n}$ is
large. (See \cite{TaoMa2007IUMM} for why $N_{n}$ needs to be large.)
But now we can not guarantee Condition $1$. Is there at least one
irrational number $\frac{\tau'}{2\pi}$ that satisfies Condition $1$
and $2$? The answer to the question is not a priori true.

In \cite{TaoMa2007IUMM}, we construct one irrational
$\frac{\tau}{2\pi}$ with delocalization. Our construction depends on
one assumption and three facts. We assume the rational case has a
nearly constant diffusion speed $v_{\tau}$ from $t=0$ to
$t\rightarrow \infty$. At least the rational case diffuses with a
nearly constant speed $v_{\tau}$ after a threshold time $t_{LD}$.
This is the momentum linear diffusion assumption, which is different
from classical diffusion. First, there is an upper limit of momentum
diffusion speed. Second, in all $\frac{\tau}{2\pi}=\frac{p}{q}$ with
different $p$, the case $\frac{\tau}{2\pi}=\frac{1}{q}$ has the
lowest diffusion speed. Third, Eq. $(21)$ connects rational and
irrational cases. The constructed $\frac{\tau}{2\pi}$ is not a
general Liouville number. To construct more irrational
$\frac{\tau}{2\pi}$, we have to exactly know the diffusion speeds of
QKR with every rational $\frac{\tau}{2\pi}$. Because some
$\frac{\tau}{2\pi}=\frac{p}{q}$ are close to $\frac{p'}{q'}$ with
$q'\ll q$, the diffusion speed $v_{\tau}$ can be far faster than
other $\frac{\tau}{2\pi}=\frac{p''}{q}$, where $\frac{p''}{q}$ is
not close to a rational number with smaller denominators. This can
be seen from Eq. $(21)$ before the divergence time
$\frac{\tau}{2\pi}=\frac{p}{q}$ and $\frac{p'}{q'}$ will not depart
from each other very much. This is one reason why the diffusion
speed $v_{\tau}$ and $v_{\tau'}$ is not significantly different from
each other. It is difficult to estimate the diffusion speed of the
general $\frac{\tau}{2\pi}=\frac{p}{q}$.

\subsection{\label{sec:level2}Some problems}
For all rational numbers $\frac{\tau}{2\pi}=\frac{p}{q}$ between 0
and 2, 0 or 2 has the fastest diffusion speed $v_{0}=k$. For
$\frac{\tau}{2\pi}=0,2$, $U(N)=F^N=e^{-i N k V(\theta)}$. For
$\frac{\tau}{2\pi}=1$, the QKR is periodic \cite{Casati1979} which
is proved in \cite{TaoMa2007Toeplitz}. How to sort all the rational
numbers according to their respective diffusion speeds? This
sequence is helpful to understand QKR. Is this ordering dependent on
$k$?

For $\tau$ close to strong resonances such as
$\{\frac{1}{3},\frac{1}{4},\frac{1}{5} \}\times 4\pi$, QKR will
diffuse quickly before the time $(\frac{6 \epsilon}{k^{2} \delta
\tau})^{1/3}$. This is the reason of anomalous localization. Even if
the majority of irrational numbers (we assume the Lebesgue measure
is 1) are localized, irrational numbers with typical localization
length can not have Lesbegue measure 1. All the irrational numbers
between, for example, $[4\pi/3-2\epsilon, 4\pi/3+2\epsilon]$, where
$\epsilon$ is a small number, do not have the typical localization
length. From Eq. $(26)$,
\begin{equation}
\frac{1}{2\epsilon} \int_{4\pi /3}^{4\pi /3+2\epsilon } l_{\tau'} \,
d\tau' =\frac{3}{2} l_{\tau'=4\pi/3+2\epsilon}.
\end{equation}
If $l_{\tau'=4\pi/3+2\epsilon}$ is the typical localization length,
the average localization length between $[4\pi/3, 4\pi/3+2\epsilon]$
is apparently not the typical localization length. Is the average
localization length $\overline{l}=\frac{1}{4\pi} \int_0^{4\pi}
l_{\tau} \, d\tau$ finite? Even if it is finite, it is not the usual
localization length $\alpha D$. In Eq. $29$, the small domain
contributes the factor $\frac{3}{2}$. $\overline{l}$ is larger than
$\alpha D$ because of the contribution from the small domain around
$\tau=4\pi /3$. Does the $\frac{1}{4\pi} \int_0^{4\pi} l_{\tau} \,
d\tau$ also contribute a factor $\frac{3}{2}$?
$\overline{l}=\frac{3}{2} \alpha D$?

The average localization length $\overline{l}$ is very difficult to
calculate. The question can be answered only after we have totally
understood delocalization of all the rational $\frac{\tau}{2\pi}$.
Surely if we did so, we would totally understand QKR. We guess the
average localization length could be infinity. Is $\int_{\tau
_1}^{\tau _2} l_{\tau} \, d\tau / |\tau_1 - \tau_2|$ finite, where
$\tau_1$ and $\tau_2$ are boundaries of a very small domain? We
guess it could be also infinity. $\overline{l}$ may be $\infty$, but
the Lesbegue measure of finite $l$ is still almost $1$. The
probability distribution of $P(l)$ is surely interesting. Our guess
is inspired by Anderson. Anderson emphasized in disordered matters
the average localization length may be infinite, nevertheless the
Lesbegue measure of finite localization length is still large. It is
the $P(l)$ rather than $\overline{l}$ that are really relevant
\cite{Anderson1978}. But the problem whether $\overline{l}$ of QKR
is finite is still interesting.

Although dynamic localization happens to irrational numbers, the key
to understand localization rests on understanding delocalization of
rational numbers, especially the delocalization speeds. In the
paper, we used the top-down method from rational cases to irrational
cases to understand irrational cases. Now we outline the bottom-up
method from the irrational cases to understand rational cases. We
use Eq. $(21)$ in reverse. For a typical irrational number with
localization, $|\frac{\tau}{2\pi}-\frac{p}{q}|<\frac{c}{q^2}$, where
is $c$ is a trivial constant, according to Dirichlet's approximation
theorem on diophantine approximation. So
$\frac{\tau}{2\pi}=\frac{p}{q}$ is still localized before the time
approximately $q^{2/3}$. If we want stronger result, we have to
prove there is a typical localized irrational $\frac{\tau}{2\pi}$,
$|\frac{\tau}{2\pi}-\frac{p}{q}|<\frac{c}{q^n}$. In this way we may
increase the time to approximately $q^{n/3}$. This is petitio
principii. But the essence is both rational and irrational numbers
are dense within each other. If typical irrational number with
typical localization length has a Lebesgue measure approximately 1,
why is there not one such number in
$[\frac{p}{q}-\frac{c}{q^n},\frac{p}{q}+\frac{c}{q^n}]$?

In the finite time of an experiment in a laboratory,
$\frac{\tau}{4\pi}=\frac{1}{3}+\frac{\sqrt{5}-1}{2 \times 10^{m}}$,
where $m$ is large, can diffuse much faster than an rational
$\frac{\tau}{4\pi}=\frac{p}{q}$, which has a large denominator $q$
and $\frac{p}{q}$ is far from any strong resonance. Another
interesting case is when $\tau$ close to $0$ or $4\pi$. Before the
diverge time between $\tau$ and 0 (or $4\pi$), QKR will diffuse with
a large speed. This is discussed from classical perspective in
\cite{Sadgrove2005}. For small time $t$, the energy grows
quadratically. When $t$ is large and not too large, the energy
growth will stop because of big denominator of $\frac{\tau}{2\pi}$
or $2-\frac{\tau}{2\pi}$ \cite{TaoMa2007IUMM}. This is referred as
dynamical freezing in \cite{Sadgrove2005}. It is difficult to
estimate the energy linear growth \cite{Sadgrove2005} at the
intermediate time from our discussion here.

\section{\label{sec:level1}Conclusions}
In the paper, we discuss some basic problems of QKR theory. We point
the flaws of the proof of dynamic localization in Fishman \textit{et
al}'s method \cite{Fishman1982}. We emphasize the physical mechanism
of Anderson localization can not totally explain dynamic
localization of QKR. We emphasize it is necessary to understand the
delocalization of all the rational $\frac{\tau}{2\pi}$s. In
\cite{TaoMa2007IUMM}, we have numerically calculated the
delocalization of $\tau=\frac{\pi}{10}$. Yes, it delocalizes. But
the delocalization time is around $2^{30}\tau$ or $2^{45}\tau$. In
\cite{TaoMa2007IUMM}, we constructed an irrational
$\frac{\tau}{2\pi}$ with delocalization. In the paper, We
theoretically prove anomalous localization and numerically confirm
it. These three phenomena tell us the QKR theory is not just
localization with irrational $\frac{\tau}{2\pi}$ and delocalization
with rational $\frac{\tau}{2\pi}$. The whole picture is more
complete only if when we take these three facts into consideration.
What we have touched is only a fraction of the QKR theory. We point
out the open problems and hope the readers can solve them.
\begin{acknowledgements}
I would like to thank Professor S. Fishman for helpful discussions.
I also thank Dr. J. Gong for pointing out their work to me. This
work is impossible to finish without my teacher Shumin Li's support.
\end{acknowledgements}

\vspace*{-.5cm}
\bibliography{QKRGeneraltheoryISubmit}

\begin{thebibliography}{27}
\expandafter\ifx\csname natexlab\endcsname\relax\def\natexlab#1{#1}\fi
\expandafter\ifx\csname bibnamefont\endcsname\relax
  \def\bibnamefont#1{#1}\fi
\expandafter\ifx\csname bibfnamefont\endcsname\relax
  \def\bibfnamefont#1{#1}\fi
\expandafter\ifx\csname citenamefont\endcsname\relax
  \def\citenamefont#1{#1}\fi
\expandafter\ifx\csname url\endcsname\relax
  \def\url#1{\texttt{#1}}\fi
\expandafter\ifx\csname urlprefix\endcsname\relax\def\urlprefix{URL }\fi
\providecommand{\bibinfo}[2]{#2}
\providecommand{\eprint}[2][]{\url{#2}}

\bibitem[{\citenamefont{Casati et~al.}(1979)\citenamefont{Casati, Chirikov,
  Izraelev, and Ford}}]{Casati1979}
\bibinfo{author}{\bibfnamefont{G.}~\bibnamefont{Casati}},
  \bibinfo{author}{\bibfnamefont{B.~V.} \bibnamefont{Chirikov}},
  \bibinfo{author}{\bibfnamefont{F.~M.} \bibnamefont{Izraelev}},
  \bibnamefont{and} \bibinfo{author}{\bibfnamefont{J.}~\bibnamefont{Ford}},
  vol.~\bibinfo{volume}{93} of \emph{\bibinfo{series}{Lecture Notes in
  Physics}} (\bibinfo{publisher}{Springer, Berlin}, \bibinfo{year}{1979}).

\bibitem[{\citenamefont{Fishman et~al.}(1982)\citenamefont{Fishman, Grempel,
  and Prange}}]{Fishman1982}
\bibinfo{author}{\bibfnamefont{S.}~\bibnamefont{Fishman}},
  \bibinfo{author}{\bibfnamefont{D.~R.} \bibnamefont{Grempel}},
  \bibnamefont{and} \bibinfo{author}{\bibfnamefont{R.~E.}
  \bibnamefont{Prange}}, \bibinfo{journal}{Phys. Rev. Lett.}
  \textbf{\bibinfo{volume}{49}}, \bibinfo{pages}{509} (\bibinfo{year}{1982}).

\bibitem[{\citenamefont{Grempel et~al.}(1982)\citenamefont{Grempel, Fishman,
  and Prange}}]{Grempel1982}
\bibinfo{author}{\bibfnamefont{D.~R.} \bibnamefont{Grempel}},
  \bibinfo{author}{\bibfnamefont{S.}~\bibnamefont{Fishman}}, \bibnamefont{and}
  \bibinfo{author}{\bibfnamefont{R.~E.} \bibnamefont{Prange}},
  \bibinfo{journal}{Phys. Rev. Lett.} \textbf{\bibinfo{volume}{49}},
  \bibinfo{pages}{833} (\bibinfo{year}{1982}).

\bibitem[{\citenamefont{Ma}({\natexlab{a}})}]{TaoMa2007Toeplitz}
\bibinfo{author}{\bibfnamefont{T.}~\bibnamefont{Ma}},
  \eprint{quant-ph/0709.2494}.

\bibitem[{\citenamefont{Moore et~al.}(1994)\citenamefont{Moore, Robinson,
  Bharucha, Williams, and Raizen}}]{Moore1994}
\bibinfo{author}{\bibfnamefont{F.~L.} \bibnamefont{Moore}},
  \bibinfo{author}{\bibfnamefont{J.~C.} \bibnamefont{Robinson}},
  \bibinfo{author}{\bibfnamefont{C.}~\bibnamefont{Bharucha}},
  \bibinfo{author}{\bibfnamefont{P.~E.} \bibnamefont{Williams}},
  \bibnamefont{and} \bibinfo{author}{\bibfnamefont{M.~G.}
  \bibnamefont{Raizen}}, \bibinfo{journal}{Phys. Rev. Lett.}
  \textbf{\bibinfo{volume}{73}}, \bibinfo{pages}{2974} (\bibinfo{year}{1994}).

\bibitem[{\citenamefont{Moore et~al.}(1995)\citenamefont{Moore, Robinson,
  Bharucha, Sundaram, and Raizen}}]{Moore1995}
\bibinfo{author}{\bibfnamefont{F.~L.} \bibnamefont{Moore}},
  \bibinfo{author}{\bibfnamefont{J.~C.} \bibnamefont{Robinson}},
  \bibinfo{author}{\bibfnamefont{C.}~\bibnamefont{Bharucha}},
  \bibinfo{author}{\bibfnamefont{B.}~\bibnamefont{Sundaram}}, \bibnamefont{and}
  \bibinfo{author}{\bibfnamefont{M.~G.} \bibnamefont{Raizen}},
  \bibinfo{journal}{Phys. Rev. Lett.} \textbf{\bibinfo{volume}{75}},
  \bibinfo{pages}{4598} (\bibinfo{year}{1995}).

\bibitem[{\citenamefont{Gong and Brumer}(2007)}]{Gong20071}
\bibinfo{author}{\bibfnamefont{J.}~\bibnamefont{Gong}} \bibnamefont{and}
  \bibinfo{author}{\bibfnamefont{P.}~\bibnamefont{Brumer}},
  \bibinfo{journal}{Phys. Rev. A} \textbf{\bibinfo{volume}{75}},
  \bibinfo{pages}{032331} (\bibinfo{year}{2007}).

\bibitem[{\citenamefont{Gong and Wang}(2007)}]{Gong20072}
\bibinfo{author}{\bibfnamefont{J.}~\bibnamefont{Gong}} \bibnamefont{and}
  \bibinfo{author}{\bibfnamefont{J.}~\bibnamefont{Wang}},
  \bibinfo{journal}{Phys. Rev. E} \textbf{\bibinfo{volume}{76}},
  \bibinfo{pages}{036217} (\bibinfo{year}{2007}).

\bibitem[{\citenamefont{Milek and Seba}(1990)}]{Milek1989}
\bibinfo{author}{\bibfnamefont{B.}~\bibnamefont{Milek}} \bibnamefont{and}
  \bibinfo{author}{\bibfnamefont{P.}~\bibnamefont{Seba}},
  \bibinfo{journal}{Phys. Rev. A} \textbf{\bibinfo{volume}{42}},
  \bibinfo{pages}{3213} (\bibinfo{year}{1990}).

\bibitem[{\citenamefont{Jitomirskaya and Simon}(1994)}]{Jitomirskaya1994}
\bibinfo{author}{\bibfnamefont{S.}~\bibnamefont{Jitomirskaya}}
  \bibnamefont{and} \bibinfo{author}{\bibfnamefont{B.}~\bibnamefont{Simon}},
  \bibinfo{journal}{Commun. Math. Phys} \textbf{\bibinfo{volume}{165}},
  \bibinfo{pages}{201} (\bibinfo{year}{1994}).

\bibitem[{\citenamefont{Anderson}(1958)}]{Anderson1958}
\bibinfo{author}{\bibfnamefont{P.~W.} \bibnamefont{Anderson}},
  \bibinfo{journal}{Phys. Rev.} \textbf{\bibinfo{volume}{109}},
  \bibinfo{pages}{1492} (\bibinfo{year}{1958}).

\bibitem[{\citenamefont{Ma}({\natexlab{b}})}]{TaoMa2007QKPR}
\bibinfo{author}{\bibfnamefont{T.}~\bibnamefont{Ma}}, \eprint{nlin/0709.2735}.

\bibitem[{\citenamefont{Bohigas et~al.}(1984)\citenamefont{Bohigas, Giannoni,
  and Schmit}}]{BGS1984}
\bibinfo{author}{\bibfnamefont{O.}~\bibnamefont{Bohigas}},
  \bibinfo{author}{\bibfnamefont{M.~J.} \bibnamefont{Giannoni}},
  \bibnamefont{and} \bibinfo{author}{\bibfnamefont{C.}~\bibnamefont{Schmit}},
  \bibinfo{journal}{Phys. Rev. Lett} \textbf{\bibinfo{volume}{52}},
  \bibinfo{pages}{1} (\bibinfo{year}{1984}).

\bibitem[{\citenamefont{Casati and Guarneri}(1984)}]{Casati1984}
\bibinfo{author}{\bibfnamefont{G.}~\bibnamefont{Casati}} \bibnamefont{and}
  \bibinfo{author}{\bibfnamefont{I.}~\bibnamefont{Guarneri}},
  \bibinfo{journal}{Commun. Math. Phys.} \textbf{\bibinfo{volume}{95}},
  \bibinfo{pages}{121} (\bibinfo{year}{1984}).

\bibitem[{\citenamefont{Casati et~al.}(1998)\citenamefont{Casati, Izrailev, and
  Sokolov}}]{Casati1998}
\bibinfo{author}{\bibfnamefont{G.}~\bibnamefont{Casati}},
  \bibinfo{author}{\bibfnamefont{F.~M.} \bibnamefont{Izrailev}},
  \bibnamefont{and} \bibinfo{author}{\bibfnamefont{V.~V.}
  \bibnamefont{Sokolov}}, \bibinfo{journal}{Phys. Rev. Lett.}
  \textbf{\bibinfo{volume}{80}}, \bibinfo{pages}{640} (\bibinfo{year}{1998}).

\bibitem[{\citenamefont{Ma}({\natexlab{c}})}]{TaoMa2007IUMM}
\bibinfo{author}{\bibfnamefont{T.}~\bibnamefont{Ma}}, \eprint{nlin/0709.2395}.

\bibitem[{\citenamefont{Lee and Ramakrishnan}(1985)}]{Lee1985}
\bibinfo{author}{\bibfnamefont{P.~A.} \bibnamefont{Lee}} \bibnamefont{and}
  \bibinfo{author}{\bibfnamefont{T.~V.} \bibnamefont{Ramakrishnan}},
  \bibinfo{journal}{Rev. Mod. Phys.} \textbf{\bibinfo{volume}{57}},
  \bibinfo{pages}{287} (\bibinfo{year}{1985}).

\bibitem[{\citenamefont{Chirikov et~al.}(1981)\citenamefont{Chirikov, Izrailev,
  and Shepelyansky}}]{Chirikov1981}
\bibinfo{author}{\bibfnamefont{B.~V.} \bibnamefont{Chirikov}},
  \bibinfo{author}{\bibfnamefont{F.~M.} \bibnamefont{Izrailev}},
  \bibnamefont{and} \bibinfo{author}{\bibfnamefont{D.~L.}
  \bibnamefont{Shepelyansky}}, \bibinfo{journal}{Sov. Sci. Rev. C}
  \textbf{\bibinfo{volume}{2}}, \bibinfo{pages}{209} (\bibinfo{year}{1981}).

\bibitem[{\citenamefont{Chirikov et~al.}(1988)\citenamefont{Chirikov, Izrailev,
  and Shepelyansky}}]{Chirikov1988}
\bibinfo{author}{\bibfnamefont{B.~V.} \bibnamefont{Chirikov}},
  \bibinfo{author}{\bibfnamefont{F.~M.} \bibnamefont{Izrailev}},
  \bibnamefont{and} \bibinfo{author}{\bibfnamefont{D.~L.}
  \bibnamefont{Shepelyansky}}, \bibinfo{journal}{Physica D}
  \textbf{\bibinfo{volume}{33}}, \bibinfo{pages}{77} (\bibinfo{year}{1988}).

\bibitem[{\citenamefont{Shepelyansky}(1986)}]{Shepelyansky1986}
\bibinfo{author}{\bibfnamefont{D.~L.} \bibnamefont{Shepelyansky}},
  \bibinfo{journal}{Phys. Rev. Lett.} \textbf{\bibinfo{volume}{56}},
  \bibinfo{pages}{677} (\bibinfo{year}{1986}).

\bibitem[{\citenamefont{Fishman et~al.}(1989)\citenamefont{Fishman, Prange, and
  Griniasty}}]{Fishman1989}
\bibinfo{author}{\bibfnamefont{S.}~\bibnamefont{Fishman}},
  \bibinfo{author}{\bibfnamefont{R.~E.} \bibnamefont{Prange}},
  \bibnamefont{and}
  \bibinfo{author}{\bibfnamefont{M.}~\bibnamefont{Griniasty}},
  \bibinfo{journal}{Phys. Rev. Lett.} \textbf{\bibinfo{volume}{39}},
  \bibinfo{pages}{1628} (\bibinfo{year}{1989}).

\bibitem[{\citenamefont{Shallit}(1982)}]{Shallit1982}
\bibinfo{author}{\bibfnamefont{J.~O.} \bibnamefont{Shallit}},
  \bibinfo{journal}{J. Number Theory} \textbf{\bibinfo{volume}{2}},
  \bibinfo{pages}{228} (\bibinfo{year}{1982}).

\bibitem[{\citenamefont{Peres}(1984)}]{Peres1984}
\bibinfo{author}{\bibfnamefont{A.}~\bibnamefont{Peres}},
  \bibinfo{journal}{Phys. Rev. A} \textbf{\bibinfo{volume}{30}},
  \bibinfo{pages}{1610} (\bibinfo{year}{1984}).

\bibitem[{\citenamefont{Berman et~al.}(1991)\citenamefont{Berman, Rubaev, and
  Zaslavsky}}]{Berman1991}
\bibinfo{author}{\bibfnamefont{G.~P.} \bibnamefont{Berman}},
  \bibinfo{author}{\bibfnamefont{V.~Y.} \bibnamefont{Rubaev}},
  \bibnamefont{and} \bibinfo{author}{\bibfnamefont{G.~M.}
  \bibnamefont{Zaslavsky}}, \bibinfo{journal}{Nonlinearity}
  \textbf{\bibinfo{volume}{4}}, \bibinfo{pages}{543} (\bibinfo{year}{1991}).

\bibitem[{\citenamefont{Casati et~al.}(1990)\citenamefont{Casati, Molinari, and
  Izrailev}}]{Casati1990}
\bibinfo{author}{\bibfnamefont{G.}~\bibnamefont{Casati}},
  \bibinfo{author}{\bibfnamefont{L.}~\bibnamefont{Molinari}}, \bibnamefont{and}
  \bibinfo{author}{\bibfnamefont{F.}~\bibnamefont{Izrailev}},
  \bibinfo{journal}{Phys. Rev. Lett.} \textbf{\bibinfo{volume}{64}},
  \bibinfo{pages}{1851} (\bibinfo{year}{1990}).

\bibitem[{\citenamefont{Anderson}(1978)}]{Anderson1978}
\bibinfo{author}{\bibfnamefont{P.~W.} \bibnamefont{Anderson}},
  \bibinfo{journal}{Rev. Mod. Phys.} \textbf{\bibinfo{volume}{50}},
  \bibinfo{pages}{191} (\bibinfo{year}{1978}).

\bibitem[{\citenamefont{Sadgrove et~al.}(2005)\citenamefont{Sadgrove,
  Wimberger, Parkins, and Leonhardt}}]{Sadgrove2005}
\bibinfo{author}{\bibfnamefont{M.}~\bibnamefont{Sadgrove}},
  \bibinfo{author}{\bibfnamefont{S.}~\bibnamefont{Wimberger}},
  \bibinfo{author}{\bibfnamefont{S.}~\bibnamefont{Parkins}}, \bibnamefont{and}
  \bibinfo{author}{\bibfnamefont{R.}~\bibnamefont{Leonhardt}},
  \bibinfo{journal}{Physical Review Letters} \textbf{\bibinfo{volume}{94}},
  \bibinfo{pages}{174103} (\bibinfo{year}{2005}).

\end{thebibliography}
\vspace*{-.5cm}
\end{document}